\documentclass[final]{elsarticle}

\usepackage{inputenc}
\usepackage{amsfonts}
\usepackage{amsthm}
\usepackage{amsmath}
\usepackage{graphicx}
\usepackage{lineno}
\usepackage{caption}
\usepackage{subfigure}
\usepackage{enumitem}

\usepackage[usenames,dvipsnames,svgnames,table]{xcolor}

\usepackage{tikz}
\usepackage{tikz-cd}
\usepackage{scalerel,stackengine}
\usepackage{pifont}
\usepackage{algorithm} 
\usepackage{algpseudocode} 
\usepackage{xspace}
\usetikzlibrary{matrix, trees, arrows}

\newif\ifshowtodo
\showtodofalse

\newif\ifshowtable
\showtablefalse

\newif\ifshowitemize
\showitemizefalse

\newcommand{\type}[1]{\texttt{#1}}

\makeatletter
\algnewcommand\MultiLineState{%
  \Statex \hspace{\algorithmicindent}\hspace{\ALG@thistlm}%
}
\makeatother

\newcommand{\R}[1]{\ensuremath{\mathbb{R}^{#1}}}

\newcommand{\N}[1]{\ensuremath{\mathbb{N}^{#1}}}

\newtheorem{theorem}{Theorem}[section]

\theoremstyle{definition}

\theoremstyle{theorem}

\theoremstyle{remark}

\newcommand{\refinementSet}[1]{\ensuremath{\mathcal{S}_{#1}}}

\newcommand{\stageOneTreeAlgorithm}{\ensuremath{\textnormal{\texttt{stageOneTree}}}}

\newcommand{\tagMaubach}[1]{\ensuremath{d_{#1}}}
\renewcommand{\tag}[1]{\ensuremath{d_{#1}}}
\newcommand{\level}[1]{\ensuremath{l_{#1}}}

\newcommand{\bisectToMaubachAlgorithm}{\ensuremath{\textnormal{\texttt{bisectCastToMaubach}}}}

\newcommand{\mesh}[1]{\ensuremath{\mathcal{T}_{#1}}}
\newcommand{\markedMesh}[1]{\ensuremath{\mesh{#1}^{\prime}}}
\newcommand{\meshQ}[2]{\ensuremath{\mathcal{Q}_{#1}^{#2}}}

\newcommand{\simplex}[1]{\ensuremath{\sigma_{#1}}}
\newcommand{\markedSimplex}[1]{\ensuremath{\rho_{#1}}}
\newcommand{\treeSimplex}[1]{\ensuremath{\tau_{#1}}}
\newcommand{\maubachSimplex}[1]{\ensuremath{\mu_{#1}}}

\newcommand{\reflectedSimplex}[1]{\ensuremath{\bar{\sigma}_{#1}}}

\newcommand{\face}[1]{\ensuremath{\kappa_{#1}}\xspace}
\newcommand{\reflectedFace}[1]{\ensuremath{\bar{\kappa}_{#1}}}

\newcommand{\vertex}[1]{\ensuremath{[v_{#1}]}}
\newcommand{\wertex}[1]{\ensuremath{[w_{#1}]}}
\newcommand{\bvertex}[1]{\ensuremath{\mathbf{v}_{#1}}}
\newcommand{\bwertex}[1]{\ensuremath{\mathbf{w}_{#1}}}
\newcommand{\multivertex}[2]{\ensuremath{[v_{#1},v_{#2}]}}

\newcommand{\edgeVertices}[2]{\ensuremath{(\vertex{#1},\vertex{#2}})}

\newcommand{\edge}[1]{\ensuremath{e_{#1}}}
\newcommand{\bisectionEdge}[1]{\ensuremath{e_{#1}}}

\newcommand{\bisectionTree}[1]{\ensuremath{t_{#1}}}

\providecommand{\etal}{\emph{et al.}}

\newcommand{\closedBall}[1]{\ensuremath{\bar{B}^{#1}}}

\newcommand{\facesDict}[1]{\ensuremath{\mathcal{D}_{#1}}}
\newcommand{\innerDict}[1]{\ensuremath{\mathcal{I}_{#1}}}
\newcommand{\boundaryDict}[1]{\ensuremath{\mathcal{B}_{#1}}}

\newcommand{\norm}[1]{\ensuremath{\left\lVert#1\right\rVert}}

\DeclareMathOperator{\tr}{tr}

\usepackage[subfigure]{tocloft}
\usepackage{xifthen}
\usepackage{wasysym}

\newcounter{exampleCounter}
\refstepcounter{exampleCounter}

\renewcommand{\vec}[1]{\mathbf{#1}}

\providecommand{\url}[1]{\texttt{#1}}

\newcounter{todoListCounter}

\newcommand{\theTodoListCounter}
{
	\arabic{todoListCounter}
}

\newcommand{\notDone}
{\textcolor{black}{$\Square$}}

\newcommand{\task}[1][\notDone]{\item[\Large #1] }

\newcommand{\listTodoName}
{ \ifthenelse{\theTodoListCounter >1}
	{
		There are \textcolor{Sepia}{\theTodoListCounter} todo's
	}
	{
		There is \textcolor{Sepia}{\theTodoListCounter} todo
	}
}

\newlistof{todo}{tmp}{\listTodoName}

\newcommand{\todo}[2]
{
	\refstepcounter{todo}
	\noindent \rule{\textwidth}{1pt} \\
	\textbf{\large \theTodoListCounter} TODO: {\sc #2} \hfill \raisebox{-1ex}{\Huge #1}\\
	\rule{\textwidth}{1pt} \\
	\vspace{-1\baselineskip}
	\addcontentsline{tmp}{todo}{{\large #1} TODO \numberline{\thetodo:}\textcolor{black}{#2}}
}

\newcommand{\listOfTodo}
{
	\ifthenelse{\theTodoListCounter > 0}
	{
		\clearpage
		\listoftodo
	}
	{
	}
}

\graphicspath{{./figures/}}

\biboptions{sort&compress}

\journal{Computer-Aided Design}

\begin{document}
	
\begin{frontmatter}
	
	
	\title{Conformal marked bisection for local refinement of \MakeLowercase{$n$}-dimensional unstructured simplicial meshes}
	
	\author[bscaddress]{Guillem Belda-Ferr{\'i}n}
	\ead{guillem.belda@bsc.es}
	
	\author[bscaddress]{Eloi Ruiz-Giron{\'e}s}
	\ead{eloi.ruizgirones@bsc.es}
	
	\author[bscaddress]{Abel Gargallo-Peir{\'o}}
	\ead{abel.gargallo@bsc.es}
	
	\author[bscaddress]{Xevi Roca\corref{mycorrespondingauthor}}
	\ead{xevi.roca@bsc.es}
	
	\cortext[mycorrespondingauthor]{Corresponding author: Xevi Roca (xevi.roca@bsc.es)}
	
	\address[bscaddress]{
		Computer Applications in Science and Engineering,\\
		Barcelona Supercomputing Center - BSC, 08034 Barcelona, Spain}
	
	\begin{abstract}
		
We present an $n$-dimensional marked bisection method for unstructured conformal meshes. We devise the method for local refinement in adaptive $n$-dimensional applications. To this end, we propose a mesh marking pre-process and three marked bisection stages. The pre-process marks the initial mesh conformingly. Then, in the first $n-1$ bisections, the method accumulates in reverse order a list of new vertices. In the second stage, the $n$-th bisection, the method uses the reversed list to cast the bisected simplices as reflected simplices, a simplex type suitable for newest vertex bisection. In the final stage, beyond the $n$-th bisection, the method switches to newest vertex bisection. To allow this switch, after the second stage, we check that under uniform bisection the mesh simplices are conformal and reflected. These conditions are sufficient to use newest vertex bisection, a bisection scheme guaranteeing key advantages for local refinement. Finally, the results show that the proposed bisection is well-suited for local refinement of unstructured conformal meshes.
		
	\end{abstract}
	
	
\end{frontmatter}


\section{Introduction}
In many applications, computational scientists and engineers need to numerically solve problems formulated in more than three dimensions, problems such as partial differential equations (PDEs) and combinatorial fixed points, problems that model physical and economic phenomena. In these problems, the dimensionality might be relatively small or potentially large. It is relatively small in problems such as 4D general relativity equations, 4D space-time discretizations of unsteady 3D phenomena, and 6D Boltzmann equations. It is potentially large in problems such as the Black-Scholes equation with as many dimensions as financial assets, the Schrödinger equation with as many dimensions as three times the number of quantum particles, and combinatorial fixed point models featuring up to thousands of dimensions. Unfortunately, the cost of solving these problems grows exponentially with dimensionality.

To mitigate this exponential growth, practitioners can use adaptive mesh refinement to reduce the element count and, thus, the computational cost. Specifically, adaptive mesh refinement uses finer elements where the solution presents larger variations and coarser elements where the solution presents smaller variations.

In adaptive $n$-dimensional refinement, conformal simplicial meshes must be locally modified. One systematic modification for arbitrary dimensions is to bisect a set of selected simplices. This operation splits each simplex by introducing a new vertex on a previously selected refinement edge. Then, this new vertex is connected to the original vertices to define two new simplices. To ensure that the mesh is still conformal, the bisection has to select additional refinement edges on a surrounding conformal closure. 

The selection of the refinement and closure edges determine different refinement approaches for simplices.
The first dominant method for refining triangles was the red-green refinement approach \cite{bank1983some}. In this method, the red refinement quadrisects a triangle by connecting the midpoints of the edges. Then, to ensure conformity, the green refinement bisects a triangle vertex to the midpoint of the opposite edge. In adaptive cycles, the green refinements are removed to quadrisect the corresponding triangles.
Later, adaptivity practitioners considered whether they could skip the green refinement and unrefinement steps. The answer to this question was the newest vertex bisection method \cite{mitchelluni, mitchell1991adaptive}. In this bisection scheme, a triangle is refined by connecting a vertex to a midpoint on a refinement edge. Then, a set of surrounding triangles is refined to preserve mesh conformity.
An alternative approach is to select the longest edge bisection, a selection that works for triangles \cite{rivara1984mesh}, tetrahedra \cite{rivara19923}, and $n$-dimensional simplices. The longest edge bisection approach features two steps, a first step that refines the marked simplices and a second step that performs additional bisections to ensure the conformal closure. Thereafter, different newest vertex bisection algorithms for tetrahedra \cite{bansch1991local, kossaczky1994recursive} and $n$-dimensional simplices \cite{maubach1995local, traxler1997algorithm} were proposed.
Finally, the marked bisection method for tetrahedra was proposed \cite{arnold2000locally}. In this method, each face has a marked edge, and the tetrahedron refinement edge is the one marked in two faces. The resulting edge marks determine five types of marked tetrahedra, types that determine five bisection rules.

In $n$-dimensional bisection, the edge selection is commonly based on choosing either the longest edge \cite{rivara1984algorithms, rivara1991local, plaza2000local, plaza2003mesh} or the newest vertex \cite{mitchell1991adaptive, kossaczky1994recursive, maubach1995local, maubach1996efficient, traxler1997algorithm}. Although both edge selections are well-suited for adaption, newest vertex bisection has been preferred in applications where it is possible to start with a reflected mesh \cite{maubach1995local,traxler1997algorithm,stevenson2008completion,alkamper2018weak}. A mesh is reflected if all the internal faces appear with the same order in the two incident simplices. Thus, the resulting bisection structure is easily forced to be the same from both sides of the face.

This preference is so since on reflected meshes newest vertex bisection guarantees key advantages for $n$-simplicial adaption. The first advantage is that if the initial mesh is conformal, the refined mesh is also conformal \cite{stevenson2008completion} (conformity). Second, local refinement of a set of simplices terminates in finite time \cite{stevenson2008completion} (finiteness). Third, successive refinement leads to a fair number of simplex similarity classes \cite{maubach1996amount,traxler1997algorithm, arnold2000locally,stevenson2008completion}. Thus, the minimum mesh quality is bounded (stability). Finally, it needs a fair number of additional bisections to complete the conformal closure \cite{stevenson2008completion} (locality).

Unfortunately, practitioners only know how to exploit the advantages of pure newest vertex bisection on complex geometries for dimensions two \cite{mitchell1991adaptive} and three \cite{belda2022bisecting}. For more dimensions, there is not any known method to extract a reflection structure from an arbitrary unstructured conformal mesh \cite{maubach1995local, traxler1997algorithm, stevenson2008completion, alkamper2018weak}.

For unstructured conformal meshes, there are pre-processing methods for $n$-dimensional meshes that allow conformal finite termination of posterior newest vertex bisection \cite{stevenson2008completion,alkamper2018weak}. However, these methods do not fulfill simultaneously the similarity and the locality properties. The number of similarity classes for the method in \cite{stevenson2008completion} is not favorably bound since the pre-process starts by splitting all simplices into $(n+1)!/2$ simplices. Thus, the method can lead to $(n+1)!/2$ times more similarity classes for each simplex than pure newest vertex bisection. On the contrary, the bound on the cost of the conformal closure for the method in \cite{alkamper2018weak} might be worse than in \cite{stevenson2008completion} since the pre-process leads to weakly reflected meshes. As stated in \cite{alkamper2018weak}, this weak condition might not satisfy the strong reflection condition required for locality \cite{stevenson2008completion}, a strong condition that is possibly more restrictive than being a weakly reflected mesh.

Fortunately, a conformal and finite bisection method is adequate for adaptive analysis if it almost fulfills the similarity and locality properties. To almost fulfill these conditions, one can use existent multi-stage methods \cite{arnold2000locally, belda2018local}. These methods feature a first stage performing a specific-purpose bisection for marked simplices. This marked bisection enforces that after a few initial steps, one can switch independently on each element to another stage featuring Maubach's newest vertex bisection \cite{maubach1995local}. The number of initial bisection steps is comparable with the spatial dimension. Hence, these steps are responsible for an initial slight increase in both the total number of similarity classes and the cost of the conformal closure. Taking into account this initial increase and since the first stage fulfills the sufficient conformity and reflection conditions stated in \cite{stevenson2008completion}, the whole method is finite and conformal while almost fulfills the similarity and locality properties. However, these multi-stage methods are specifically devised for 3D \cite{arnold2000locally} and 4D \cite{belda2018local} meshes, yet they are practical bisection methods.

Accordingly, the question of whether there is a practical $n$-dimensional multistage bisection method for unstructured conformal meshes in more than four dimensions is still open.To formally answer this question is our long-term goal, yet it is out of the scope of this work. Our current goal is to heuristically propose a practical $n$-dimensional multi-stage bisection on unstructured conformal meshes.

To meet our goal, the main contribution is to propose and implement an $n$-dimensional three-stage bisection method. By construction, the method starts by marking the sub-simplices of the initial mesh conformingly. Then, independently for each simplex, the first $n-1$ marked bisections accumulate in reverse order the list of newly created mid-edge vertices (first stage). The $n$-th marked bisection completes the accumulated vertex list to replace simplices of bisection level $n$ with equivalent simplices having a Maubach tag equal to $n$ (second stage). This equivalent tagging allows switching to Maubach's newest vertex bisection (third stage). The pre-process and the three stages work in $n$ dimensions by construction, and hence, the method is $n$-dimensional. Furthermore, the method heuristically enforces conformity and finiteness while almost meeting the similarity and locality properties. Although we do not formally prove that the conformity and reflectivity conditions are satisfied, we perform the corresponding experiments to check them. We finally apply the implementation of the proposed method to successfully locally refine unstructured conformal meshes.

The rest of the paper is organized as follows. In Section \ref{sec:preliminaries}, we introduce the preliminary notation and concepts. 
Following, in Section \ref{sec:problemStatement}, we state the problem, and we outline the solution. In Section \ref{sec:definitions}, we introduce some notation and definitions specific to our method. Next, in Section \ref{sec:bisectionalgorithm}, we propose a co-dimensional marking pre-process and a three-stages marked bisection method. In Section \ref{sec:examples}, we present several examples to show the features of the proposed method. Finally, in Section \ref{sec:conclusions}, we present the conclusions and the future work.

\section{Preliminaries}
\label{sec:preliminaries}

We proceed to introduce the necessary notation and concepts. Specifically, we introduce the preliminaries related to simplicial meshes, bisection methods, and bisection trees. 

\subsection{Simplicial meshes and bisection}
\label{sec:simplicialMeshes}

A \emph{simplex} is the convex hull of $n+1$ points $p_{0},\ldots, p_{n} \in \R{n}$ that do not lie in the same hyper-plane. We denote a simplex as $\simplex{}\ = conv(p_{0},\ldots, p_{n})$. We identify each point $p_{i}$ with a unique integer identifier $v_{i}$ that we refer as \emph{vertex}. Thus, a simplex is composed of $n+1$ vertices and we denote it as $\simplex{} = (v_{0}, \ldots, v_{n})$ where $v_{i}$ is the identifier of point $p_{i}$. We have an application $\Pi$ that maps each identifier $v_{i}$ to the corresponding point as $\Pi(v_{i}) = p_{i}$. 

Given a simplex \simplex{}, a \emph{$k$-entity} is a sub-simplex composed of $k+1$ vertices of \simplex{}, for $0 \leq k \leq n-1$. We say that a 2-entity is an \emph{edge} and a $(n-1)$-entity is a \emph{face}. We associate each face of a simplex \simplex{}\ to its opposite vertex in \simplex{}. Specifically, the opposite face to $v_{i}$ is
\[
\face{i} = (v_{0}, v_{1}, \ldots, v_{i-1}, v_{i+1}, \ldots, v_{n}).\]
We say that two simplices \simplex{1}\ and \simplex{2}\ are \emph{neighbors} if they share a face.

We define a \emph{mesh}, \mesh{}, associated to an open set $\Omega \in \R{n}$ as a finite collection of mutually disjointed simplices such that
\[\overline{\Omega} = \bigcup_{\simplex{} \in \mesh{}} \simplex{}.\]
A simplicial mesh is \emph{conformal} if the following two conditions are satisfied: 
\begin{enumerate}
	\item[(C1)] For any $\simplex{} \in \mesh{}$, $\simplex{} \cap \partial \Omega$ is the union of entities of \simplex{}.
	\item[(C2)] For any $\simplex{1}, \simplex{2} \in \mesh{}$, either $\simplex{1} \cap \simplex{2}$ is empty, or a $k$-entity of \mesh{}\ with $0 \leq k \leq n-1$. 
\end{enumerate} 

\begin{figure}[t!]
	\centering
	\includegraphics[width=0.4\textwidth]{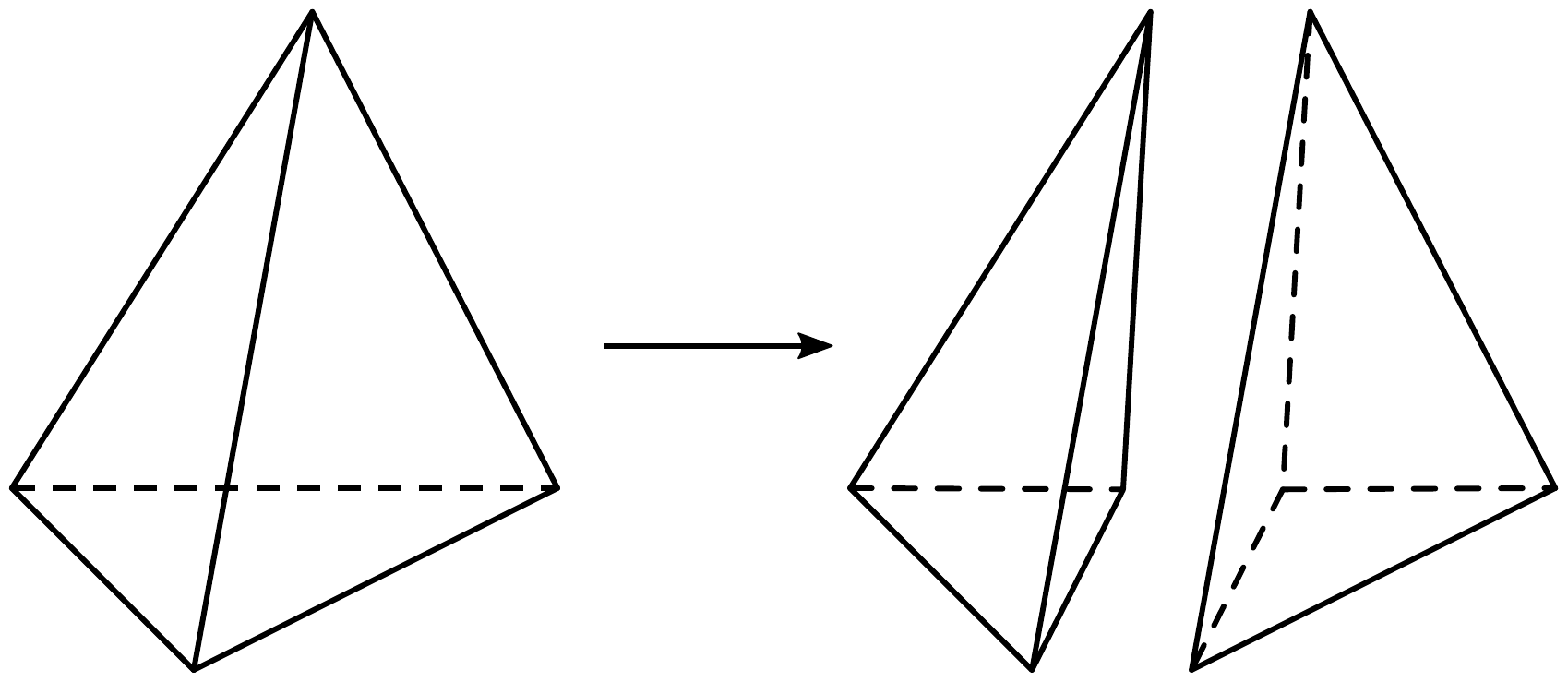}
	\caption{Bisection of a tetrahedron.}
	\label{fig:bisection}
\end{figure}

We define the \emph{bisection} of a simplex as the operation that splits a simplex by introducing a new vertex on the selected refinement edge. Then, the vertices not lying on this refinement edge are connected to the new vertex. 
These connections determine two new simplices, a first simplex determined by the refinement vertex and the $n$ vertices opposite to the second vertex of the refinement, and a second simplex determined by the refinement vertex and the $n$ vertices opposite to the first vertex of the refinement edge.
Figure \ref{fig:bisection} shows the bisection of a tetrahedron.

\subsection{Marked bisection}

Arnold \etal\ \cite{arnold2000locally} presented a marked bisection algorithm for unstructured conformal tetrahedral meshes that ensures locally refined conformal meshes and quality stability. 
The \emph{refinement edge} is the edge of \simplex{}\ to be bisected. Since an edge is shared by $n-1$ faces of the simplex, the faces that contain the refinement edge are the \emph{refinement faces} of \simplex{}. 
Because the simplex has $n+1$ faces, only two faces remain. These faces are opposite to each of the vertices of the refinement edge, and they are called \emph{non-refinement faces}.

Now, we can introduce the definition of \emph{marked simplex}, which is a modification of the one detailed in \cite{arnold2000locally}. Herein, a marked simplex is a simplex \simplex{}\ equipped with a data structure that tells us how to refine it and its descendants.

A mesh is \emph{marked} if all its simplices are marked. A marked conformal mesh is \emph{conformingly marked} if each entity has a unique marked edge. That is, an entity shared by two simplices has the same marked edge from both sides. Accordingly, shared entities are bisected in the same manner from different simplices. 

\begin{algorithm}[t!]
	\caption{Refining a subset of a mesh.}
	\label{alg:RefineMeshAlgorithm}
	\begin{algorithmic}[1]
		\Require{\type{Mesh} \mesh{}, \type{SimplicesSet} $\refinementSet{} \subset \mesh{}$}
		\Ensure{\type{ConformalMarkedMesh} $\mesh{2}$}
		\Function{refineMesh}{\mesh{}, \refinementSet{}}
		\State $\mesh{1} = \Call{markMesh}{\mesh{}}$
		\State $\mesh{2} = \Call{localRefine}{\mesh{1}, \refinementSet{}}$
		\State \Return \mesh{2}
		\EndFunction
	\end{algorithmic}	
\end{algorithm}
\begin{algorithm}[t!]
	\caption{Local refinement of a marked mesh.}
	\label{alg:LocalRefineAlgorithm}
	\begin{algorithmic}[1]
		\Require{\type{ConformalMarkedMesh} \mesh{} and \type{SimplicesSet} $\refinementSet{} \subset \mesh{}$}
		\Ensure{\type{ConformalMarkedMesh} $\mesh{}^{\prime}$}
		\Function{localRefine}{\mesh{}, \refinementSet{}}
		\State $\bar{\mesh{}} = \Call{bisectSimplices}{\mesh{}, \refinementSet{}}$
		\State $\mesh{}^{\prime} = \Call{refineToConformity}{\bar{\mesh{}}}$
		\State $\mesh{0} = \Call{renumberMesh}{\mesh{}^{\prime}}$
		\label{line:renumberMesh}
		\State \Return $\mesh{0}$
		\EndFunction
	\end{algorithmic}	
\end{algorithm}
\begin{algorithm}[t!]
	\caption{Refine-to-conformity a marked mesh.}
	\label{alg:RefineToConformityAlgorithm}
	\begin{algorithmic}[1]
		\Require{\type{MarkedMesh} \mesh{}}
		\Ensure{\type{MarkedMesh} $\mesh{}^{\prime}$ without hanging vertices}
		\Function{refineToConformity}{$\mesh{}\,$}
		\State $\refinementSet{} = \Call{getNonConformalSimplices}{\mesh{}}$
		\If {$\refinementSet{} \neq \emptyset$}
		\label{line:RefineToConformity_SnotEmpty}
		\State $\bar{\mesh{}} = \Call{bisectSimplices}{\mesh{}, \refinementSet{}}$
		\label{line:RefineToConformity_bisectPentatopes}
		\State $\mesh{}^{\prime} = \Call{refineToConformity}{\bar{\mesh{}}}$
		\label{line:RefineToConformity_RefineToConformity}
		\Else
		\State $\mesh{}^{\prime} = \mesh{}$
		\label{line:RefineToConformity_finish}
		\EndIf
		\State \Return $\mesh{}^{\prime}$
		\EndFunction
	\end{algorithmic}
\end{algorithm}
\begin{algorithm}[t!]
	\caption{Bisect a set of simplices.}
	\label{alg:bisectSimplices}
	\begin{algorithmic}[1]
		\Require{\type{MarkedMesh} \mesh{}, \type{SimplicesSet} \refinementSet{}}
		\Ensure{\type{MarkedMesh} \mesh{1}}
		\Function{bisectSimplices}{\mesh{}, \refinementSet{}}
		\State $\mesh{1} = \emptyset$
		\For{$\markedSimplex{} \in \mesh{}$}
		\If{$\markedSimplex{} \in \refinementSet{}$}
		\State $\markedSimplex{1}, \markedSimplex{2} = \Call{bisectSimplex}{\markedSimplex{}}$
		\State $\mesh{1} = \mesh{1} \cup \markedSimplex{1}$
		\State $\mesh{1} = \mesh{1} \cup \markedSimplex{2}$
		\Else
		\State $\mesh{1} = \mesh{1} \cup \markedSimplex{}$
		\EndIf
		\EndFor
		\State \Return \mesh{1}
		\EndFunction
	\end{algorithmic}
\end{algorithm}
To perform the bisection process, we adapt to the $n$-dimensional case the recursive refine-to-conformity scheme proposed in \cite{arnold2000locally}. The marked bisection method, Algorithm \ref{alg:RefineMeshAlgorithm}, starts by marking the initial unstructured conformal mesh and then applies a local refinement procedure to a set of simplices of the marked mesh. 
To do it so, we need to specify a conformal marking procedure for simplices to obtain a marked mesh \markedMesh{}.
Using this marked mesh, the local refinement procedure, Algorithm \ref{alg:LocalRefineAlgorithm}, first refines a set of simplices, then calls a recursive refine-to-conformity strategy, and finally renumbers the mesh, see  \ref{sec:renumber}. The refine-to-conformity strategy, Algorithm \ref{alg:RefineToConformityAlgorithm}, terminates when successive bisection leads to a conformal mesh. Both algorithms use marked bisection to refine a set of elements, see Algorithm \ref{alg:bisectSimplices}.

\subsection{Newest vertex bisection}

We introduce Maubach's bisection \cite{maubach1995local} and some definitions of Stevenson \cite{stevenson2008completion} that present the sufficient conditions to use newest vertex bisection.

\begin{algorithm}[t!]
	\caption{Maubach's bisection of a $n$-simplex.}
	\label{alg:maubach}
	\begin{algorithmic}[1]
		\Require{\type{TaggedSimplex} $\simplex{}$}
		\Ensure{\type{TaggedSimplex} $\simplex{1}$, \type{TaggedSimplex} $\simplex{2}$}
		
		\Function{bisectTaggedSimplex}{\simplex{}}
		\State $(v_{0},v_{1}, \ldots , v_{n})_{d} = \simplex{}$
		\State Set 
		$
		d^{\prime} = \left\{
		\begin{array}{ll} 
			d-1, & d > 1 \\ 
			n, & d = 1 
		\end{array}
		\right.
		$
		\label{line:setNewTag}
		\State Create the new vertex $z = \dfrac{1}{2}(v_{0} + v_{d})$
		\label{line:setNewNodeMaubach}
		\State $\simplex{1} = (v_{0}, v_{1}, \ldots, v_{d-1},z, v_{d+1}, \ldots, v_{n})_{d^{\prime}}$.
		\label{line:setNewChildre1Maubach}
		\State $\simplex{2} = (v_{1}, v_{2}, \ldots, v_{d},z, v_{d+1}, \ldots, v_{n})_{d^{\prime}}$.
		\label{line:setNewChildre2Maubach}
		\State \Return \maubachSimplex{1}, \maubachSimplex{2}
		\EndFunction
	\end{algorithmic}
\end{algorithm}
We define a \emph{tagged simplex} as a simplex $\simplex{} = (v_{0},v_{1}, \ldots , v_{n})$ equipped with an integer called \emph{tag}, $\tag{} \in \{1, \ldots, n\}$. We denote it as $\simplex{} = (v_{0},v_{1}, \ldots , v_{n})_{d}$.
 
According to Stevenson \cite{stevenson2008completion}, we say that two neighboring tagged simplices $\simplex{1} = (v_{0}, v_{1}, \ldots, v_{n})_{\tag{1}}$ and $\simplex{2} = (w_{0}, w_{1}, \ldots, w_{n})_{\tag{2}}$ with $\tag{1} = \tag{2}$ are \emph{reflected neighbors} if the order of vertices of the shared face is the same. That is, there are two sequences $i_{0} < i_{1} < \ldots < i_{n-1}$ and $j_{0} < j_{1} < \ldots < j_{n-1}$ such that $v_{i_{k}} = w_{j_{k}}$, for $k = 0, \ldots, n-1$ and $i_{k}, j_{k} \in \N{}$. Traxler \cite{traxler1997algorithm} defines a mesh as \emph{reflected} if all the adjacent pairs of simplices are reflected.

Alk\"{a}mper \etal\ \cite{alkamper2018weak} introduce some definitions about the compatibility of a face and of a mesh under bisection. A face $\face{} = \simplex{1} \cap \simplex{2}$ of the mesh \mesh{}\ is called a \emph{strongly compatible face} if \simplex{1}\ and \simplex{2} are reflected neighbors, or their children adjacent to \face{}\ are reflected neighbors. A mesh \mesh{}\ is a \emph{strongly compatible mesh} if all faces in \mesh{}\ are strongly compatible. Note that to have a strongly compatible mesh, it is sufficient to have a reflected mesh.

If the initial mesh is strongly compatible, local refinement with newest vertex bisection generates a conformal mesh and terminates in a finite number of steps, see  \cite[Theorem 5.1]{maubach1995local} and \cite[Theorem 5.1]{stevenson2008completion}. Furthermore, successive refinement of newest vertex bisection leads to $M_{n} = n n! 2^{n-2}$ similarity classes, see \cite[Theorem 4.5]{arnold2000locally}. Thus, the minimum mesh quality of the refined mesh is bounded.

\subsection{Bisection trees}
\label{sec:bisectionTrees}

During the bisection process, we select a bisection edge to perform the bisection of \simplex{}. 
The process of selecting the bisection edge is repeated for the two children, the four grandchildren, and so on.
We can encode the bisection process by storing those edges in a binary tree \cite{maubach1995local} that, at level \level{}, has the edges to be bisected of the descendants of level \level{}. 
We define this binary tree as the \emph{bisection tree} of \simplex{}, and we denote it as \bisectionTree{}. Let $\simplex{} = (v_{0}, v_{1}, \ldots, v_{n})$ be a simplex and \bisectionTree{}\ its bisection tree. The order of the edges of \bisectionTree{}\ is determined by
\begin{equation}
	\label{eq:localEdges}
	\begin{array}{*{4}{c}}
		(v_{0},v_{1}) & (v_{0}, v_{2}) &  \cdots & (v_{0},v_{n}) \\
		& (v_{1},v_{2}) & \cdots & (v_{1},v_{n}) \\
		& & \cdots & (v_{2},v_{n}) \\
		& & \ddots & \vdots \\
		& & (v_{n-2},v_{n-1}) & (v_{n-2},v_{n}) \\
		& & & (v_{n-1},v_{n}).
	\end{array}
\end{equation}

\begin{figure}[t!]
	\centering
	\begin{tabular}{cc}
		\subfigure[]{
		\begin{tikzpicture}
			[level distance=1cm,
			level 1/.style={sibling distance=1.75cm}]
			\node {$(v_{0},v_{2})$}
			child {
				node {$(v_{0},v_{1})$}
			}
			child {
				node{$(v_{1},v_{2})$}
			};
			
		\end{tikzpicture}
		\label{fig:balancedBisectionTree}
		}
		&
		\subfigure[]{
			\includegraphics[width=0.3\textwidth]{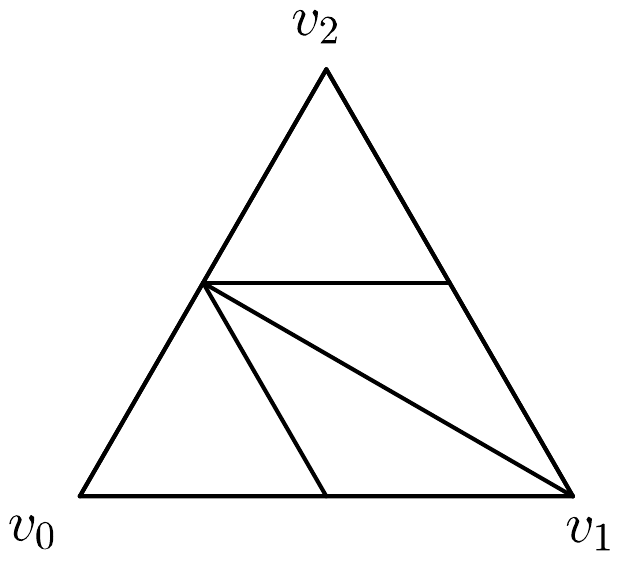}
			\label{fig:maubach2dtag2}
		}
	\end{tabular}
	\caption{\subref{fig:balancedBisectionTree} Bisection tree of the tagged triangle $\simplex{} = (v_{0},v_{1},v_{2})_{2}$ generated by Maubach's algorithm and \subref{fig:maubach2dtag2} the corresponding conformal triangular mesh after two uniform refinements.}
	\label{fig:balancedBisectionTreeTriangle}
\end{figure}
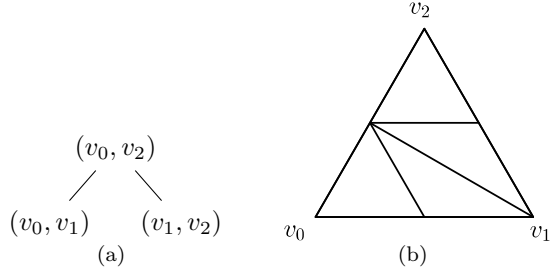

In Figure \ref{fig:balancedBisectionTreeTriangle}, we illustrate the bisection tree generated by Maubach's bisection applied to a triangle $\simplex{} = (v_{0}, v_{1}, v_{2})_{2}$. This bisection tree has height two, contains all the edges of \simplex{}, and leads to a conformal triangular mesh after two uniform refinements.

\section{Problem and outline of our solution}
\label{sec:problemStatement}

Our goal is to locally refine an $n$-dimensional conformal unstructured simplicial mesh using a stable bisection method that ensures a conformal mesh in a finite number of steps.
The input of our problem is an unstructured conformal simplicial mesh \mesh{}\ and a set of elements \refinementSet{}\ to be refined. The output is a conformal unstructured marked simplicial mesh  \mesh{1} with the simplices in \refinementSet{}\ bisected.

\begin{algorithm}[t!]
	\caption{Bisection of a marked simplex \markedSimplex{}.}
	\label{alg:bisectSimplex}
	\begin{algorithmic}[1]
		\Require{\type{MarkedSimplex} \markedSimplex{}}
		\Ensure{\type{MarkedSimplex} \markedSimplex{1}, \type{MarkedSimplex} \markedSimplex{2}}
		\Function{bisectSimplex}{\markedSimplex{}}
		\State $\level{} = \Call{level}{\markedSimplex{}}$
		\If{$\level{} < n-1$}
		\label{line:iniciStageOne}
		\State $\treeSimplex{} = \Call{TreeSimplex}{\markedSimplex{}}$
		\State $\treeSimplex{1}, \treeSimplex{2} = \Call{bisectStageOne}{\treeSimplex{}}$
		\State $\markedSimplex{1}, \markedSimplex{2} = \Call{MarkedSimplex}{\treeSimplex{1}, \treeSimplex{2}}$
		\label{line:finalStageOne}
		\ElsIf{$\level{} = n-1$}
		\label{line:iniciCast}
		\State $\treeSimplex{} = \Call{TreeSimplex}{\markedSimplex{}}$
		\State $\maubachSimplex{1}, \maubachSimplex{2} = \Call{\bisectToMaubachAlgorithm}{\treeSimplex{}}$
		\State $\markedSimplex{1}, \markedSimplex{2} = \Call{MarkedSimplex}{\maubachSimplex{1}, \maubachSimplex{2}}$
		\label{line:finalCast}
		\Else
		\label{line:iniciMaubach}
		\State $\maubachSimplex{} = \Call{MaubachSimplex}{\markedSimplex{}}$
		\State $\maubachSimplex{1}, \maubachSimplex{2} = \Call{bisectMaubach}{\maubachSimplex{}}$
		\State $\markedSimplex{1}, \markedSimplex{2} = \Call{MarkedSimplex}{\maubachSimplex{1}, \maubachSimplex{2}}$
		\label{line:finalMaubach}
		\EndIf
		\State \Return \markedSimplex{1}, \markedSimplex{2}
		\EndFunction
	\end{algorithmic}
\end{algorithm}
The kernel of our bisection method, Algorithm \ref{alg:bisectSimplex}, bisects a single simplex. This algorithm is divided into three stages determined by the descendant level of the simplex to be bisected. The first stage (Lines \ref{line:iniciStageOne}--\ref{line:finalStageOne}) and the second stage (Lines \ref{line:iniciCast}--\ref{line:finalCast}) promote that we obtain a conformal reflected mesh. In this manner, we ensure a strongly compatible mesh which is a sufficient condition to use Maubach's algorithm. The last stage is Maubach's algorithm, see Lines \ref{line:iniciMaubach}--\ref{line:finalMaubach}. Maubach's algorithm can be applied locally ensuring the generation of conformal meshes in a finite number of steps. In addition, it is the only known bisection strategy for $n$-dimensional simplicial meshes that has been proved to guarantee a finite number of similarity classes, and therefore, we ensure that our bisection strategy is stable. 

\section{Definitions in our method}
\label{sec:definitions}

Before we detail our marked bisection method, we introduce: the notion of multi-id to provide a unique identifier to the mid-vertices; the search of simplices that define a non-conformal configuration; and the selection of the bisection edge in a consistent manner.

\subsection{Unique mid-vertex identifiers: multi-ids}
\label{sec:multiids}
\begin{algorithm}[t!]
	\caption{Generation of a new multi-id.}
	\label{alg:newMultivertex}
	\begin{algorithmic}[1]
		\Require{\type{Multi-Id} \bvertex{1}, \type{Multi-Id} \bvertex{2}}
		\Ensure{\type{Multi-Id} \bvertex{}}
		\Function{midVertex}{$\bvertex{1}, \bvertex{2}$}
		\State $[v_{1,1},  \ldots , v_{1,k_{1}}] = \bvertex{1}$
		\State $[v_{2,1},  \ldots , v_{2,k_{2}}] = \bvertex{2}$
		\State $[v_{i_{1}}, \ldots, v_{i_{k_{1}+k_{2}}}] =$
		\MultiLineState $\Call{sort}{[v_{1,1}, \ldots , v_{1,k_{1}}, v_{2,1}, \ldots , v_{2,k_{2}}]$}
		\State $\bvertex{} = [v_{i_{1}}, \ldots, v_{i_{k_{1}+k_{2}}}]$
		\State \Return $\bvertex{}$
		\EndFunction
	\end{algorithmic}
\end{algorithm}

We use \emph{multi-ids} to uniquely identify the new vertices that are created during the bisection process. A multi-id is a sorted list of vertices, $\bvertex{} = [v_{1}, \ldots, v_{k}]$, where $v_{1} \leq v_{2} \leq \ldots \leq v_{k}$.    
Particularly, a vertex id $v$ can be mapped to a multi-id of length one as \vertex{}. A simplex that contains multi-ids is denoted as $\simplex{} = (\bvertex{0}, \ldots, \bvertex{n})$. We reinterpret a simplex $\simplex{}=(v_{0}, \ldots, v_{n})$ using multi-ids as $\simplex{} = (\vertex{0}, \ldots, \vertex{n})$. 

When creating a new vertex after bisecting and edge, we generate a multi-id for the new vertex. The new multi-id is the combination of the multi-ids of the edge vertices, \bvertex{0}\ and \bvertex{1}, see Algorithm \ref{alg:newMultivertex}. That is, \bvertex{}\ is the multi-id of the new vertex after bisecting $\edge{} = (\bvertex{0}, \bvertex{1})$. The resulting multi-id is created by merging and sorting the the multi-ids of \bvertex{0}\ and \bvertex{1}. We remark that the ids can appear more than once after generating a new multi-id.

We sort the multi-ids using a lexicographic order. Let $\bvertex{i} = [v_{i_{1}}, \ldots, v_{i_{k_{i}}}]$ and $\bvertex{j} = [v_{j_{1}}, \ldots, v_{j_{k_{j}}}]$ be two multi-ids, we say that $\bvertex{i} < \bvertex{j}$ if there exists $r$ such that $v_{i_{l}} < v_{j_{l}}$ for all $l < r$ and $v_{i_{l}} = v_{j_{l}}$.

\subsection{Non-conformal simplices: hanging vertices}
\begin{algorithm}[t!]
	\caption{Simplices with hanging vertices of the mesh \mesh{}.}
	\label{alg:hangingNodes}
	\begin{algorithmic}[1]
		\Require{\type{Mesh} \mesh{}}
		\Ensure{\type{SimplicesSet} \refinementSet{}}
		\Function{getNonConformalSimplices}{\mesh{}}
		\State $\refinementSet{} = \emptyset$
		\State $\mathcal{V} = \Call{Vertices}{\mesh{}}$
		\For{$\simplex{} \in \mesh{}$}
		\For{$\edge{} \in \simplex{}$}
		\State $(\bvertex{0}, \bvertex{1}) = \edge{}$
		\State $\bvertex{} = \Call{midVertex}{\bvertex{0}, \bvertex{1}}$ 
		\If{$\bvertex{} \in \mathcal{V}$}
		\State $\refinementSet{} = \refinementSet{} \cup \simplex{}$
		\EndIf
		\EndFor
		\EndFor
		\State \Return $\refinementSet{}$
		\EndFunction
	\end{algorithmic}
\end{algorithm}
In the recursive refining strategy, see Algorithm \ref{alg:RefineToConformityAlgorithm}, we need to detect and refine the non-conformal simplices of the mesh.  The non-conformal configurations in the mesh are created when refining the adjacent simplices. Thus, the main idea is to loop on the mesh simplices to detect those edges that overlap with an edge bisected from a neighboring element. This bisected edge features a mid-vertex which is seen as a hanging vertex from the edge of the current simplex. The simplices with at least one edge with a hanging vertex define the refinement set to enforce a conformal mesh.

To get the refinement set, Algorithm \ref{alg:hangingNodes} loops over all the simplices of the mesh. For each simplex, if any of its edges has a hanging mid-vertex, we add the simplex in the refinement set. To check if an edge contains a mid-vertex, we use the multi-ids. For a given edge $\edge{} = (\bvertex{0},\bvertex{1})$, we obtain the associated multi-id, $\bvertex{}$, of the mid-vertex using Algorithm \ref{alg:newMultivertex}. If the multi-id \bvertex{}\ is in the set of mesh vertices, the edge $e$ contains a hanging mid-vertex and therefore, the simplex defines a non-conformal configuration.

\subsection{Unequivocal edge selection per mesh entity: consistent bisection edge}
\label{sec:consistentBisectionEdge}

For all mesh entities shared by different mesh elements, we must ensure that these entities have the same bisection edge on all those elements. To this end, we base this selection on a strict total order of the mesh edges. The main idea is to order the edges from the longest one to the shortest one, and use a tie-breaking rule for the edges with the same length. Specifically, we define the consistent bisection edge of a simplex as the longest edge with the lowest global index.

A shared edge between two simplices may have a different order of vertices, which can induce different results when computing the edge length from different elements. To avoid these discrepancies, we use the concept of \emph{global edges}. A global edge is a pair of indices,  $\bisectionEdge{} = (\bvertex{i},\bvertex{j})$, with $\bvertex{i} < \bvertex{j}$. The length of a global edge is defined as $\norm{e} = \norm{\Pi(\bvertex{j}) - \Pi(\bvertex{i})}$, where $\norm{\cdot}$ denotes the Euclidean norm. We compute the length of an edge using the corresponding global edge. Therefore, for different views of the same global edge, we always obtain the same length since we perform the same numerical operations in the same order.

To define a strict total order of edges, when two edges have the same length, we need a tie-breaking rule. To this end, we use a lexicographic order for the global edges in terms of the order of the vertices. We say that the global edge $\edge{i} = (\bvertex{i_{1}},\bvertex{i_{2}})$ has lower global index than the global edge $\edge{j} = (\bvertex{j_{1}},\bvertex{j_{2}})$ if $\bvertex{i_{1}} < \bvertex{j_{1}}$, or $\bvertex{i_{1}} = \bvertex{j_{1}}$ and $\bvertex{i_{2}} < \bvertex{j_{2}}$. Note that the proposed lexicographic order is strict and total since it is straight-forward to check that is irreflexive, transitive, asymmetric, and connected. We identify each global edge with a unique integer.  To this end, we sort all the existing edges of the mesh using the global index criteria. Then, we sort the edges by length using a stable sorting method. This guarantees that edges with the same length maintain the order given by the lowest global index. 

The \emph{consistent bisection edge} of a simplex is the edge with the lowest integer assigned in the edge ordering process. Note that the consistent bisection edge of a simplex is unique because we use a strict total order to define it.

\section{Marked bisection method}
\label{sec:bisectionalgorithm}

We proceed to detail a new marked bisection method. Firstly, we detail the co-dimensional marking process, which is based on the consistent bisection edge of a simplex. Then, we define the three stages of the bisection process.

\subsection{Pre-processing: co-dimensional marks}
\label{sec:markingProcess}

\begin{algorithm}[t!]
	\caption{Mark a $k$-simplex.}
	\label{alg:markingProcedureSimplices}
	\begin{algorithmic}[1]
		\Require{\type{$k$-Simplex} \simplex{}}
		\Ensure{\type{BisectionTree} \bisectionTree{}}
		
		\Function{\stageOneTreeAlgorithm}{\simplex{}}
		\State \bisectionEdge{} = \Call{consistentBisectionEdge}{\simplex{}}
		\label{line:consistentBisectionEdge}
		\If{$\dim{\simplex{}} = 1$}
		\label{line:ifDimensionEqualOne}
		\State $\bisectionTree{} = \texttt{tree}(\texttt{node}=\bisectionEdge{})$
		\label{line:returnLeaf}
		\Else
		\label{line:ifDimensionNotEqualOne}
		\State $\edgeVertices{1}{2} = \bisectionEdge{}$
		\State \face{1} = \Call{oppositeFace}{\simplex{}, \vertex{1}}
		\label{line:computeOppositeFace1}
		\State \face{2} = \Call{oppositeFace}{\simplex{}, \vertex{2}}
		\label{line:computeOppositeFace2}
		\State \bisectionTree{1} = \Call{\stageOneTreeAlgorithm}{\face{1}}
		\label{line:computeBranches1}
		\State \bisectionTree{2} = \Call{\stageOneTreeAlgorithm}{\face{2}}
		\label{line:computeBranches2}
		\State $\bisectionTree{} = \texttt{tree}(\texttt{node} = \bisectionEdge{}, \texttt{left} = \bisectionTree{1}, \texttt{right} = \bisectionTree{2}) $
		\label{line:refinementTree}
		\EndIf
		\State \Return \bisectionTree{}
		\EndFunction
	\end{algorithmic}
\end{algorithm}
\begin{algorithm}[t!]
	\caption{Mark a conformal simplicial mesh.}
	\label{alg:markingProcedureMesh}
	\begin{algorithmic}[1]
		\Require{\type{ConformalMesh} \mesh{}}
		\Ensure{\type{ConformalMarkedMesh} \markedMesh{}}
		\Function{markMesh}{\mesh{} }
		\State $\markedMesh{} = \emptyset$
		\label{line:createMarkedMesh}
		\For{$\simplex{} \in \mesh{}$} 
		\label{line:loopSimplices}
		\State \bisectionTree{} = \Call{\stageOneTreeAlgorithm}{\simplex{}} 
		\label{line:computeMarksSimplex}
		\State $\reflectedFace{} = (\,)$ 
		\label{line:initializeReflectedSimplex}
		\State \level{} = 0 
		\label{line:descendantLevelZero}
		\State $\markedSimplex{} =(\simplex{}, \reflectedFace{}, \bisectionTree{}, \level{})$ 
		\label{line:setTreeSimplex}
		\State $\markedMesh{} = \markedMesh{} \cup \markedSimplex{} $
		\label{line:addMarkSimplex}
		\EndFor
		\State \Return \markedMesh{}
		\EndFunction
		\label{line:returnMarkedMesh}
	\end{algorithmic}
\end{algorithm}

We propose a co-dimensional marking process for a simplex, in which the resulting mark is a tree. 
The tree is computed by traversing the sub-entities of the simplex in a recursive manner and selecting the consistent bisection edge of each sub-simplex. The resulting \emph{bisection tree} has height $n$, and the tree nodes of level $i$ correspond to the consistent bisection edges of sub-simplices of co-dimension $i$ (dimension $n-i$). 

Next, we detail the co-dimensional marking process for a single simplex, Algorithm \ref{alg:markingProcedureSimplices}. Since the co-dimensional marking process is the first step of the mesh refinement algorithm, the length of the multi-ids of all simplices is one. The input of the function is a simplex $\simplex{} = (\vertex{0},\ldots,\vertex{n})$ and the output is the corresponding bisection tree.  First, we obtain the consistent bisection edge, \bisectionEdge{}, of the simplex, see Line \ref{line:consistentBisectionEdge}. If \simplex{}\ is an edge, this corresponds to the base case of the recursion and we return a tree with only the root node. Otherwise, we obtain the opposite faces of the vertices of the bisection edge, see Lines \ref{line:computeOppositeFace1}--\ref{line:computeOppositeFace2}. Then, we recursively call the marking process algorithm for the faces \face{1}\ and \face{2}, and we obtain the corresponding trees \bisectionTree{1}\ and \bisectionTree{2}, see Lines \ref{line:computeBranches1}--\ref{line:computeBranches2}. Finally, we build the bisection tree \bisectionTree{}\ with the bisection edge as root node and the trees \bisectionTree{1}\ and \bisectionTree{2}\ as left and right branches, see Line \ref{line:refinementTree}.

Then, we obtain a marked mesh by marking all mesh simplices, see Algorithm \ref{alg:markingProcedureMesh}. The input is a conformal simplicial mesh, \mesh{}, and the output is a conformal marked simplicial mesh, \markedMesh{}.  A \emph{tree-simplex} is a 4-tuple $\treeSimplex{} = (\simplex{}, \reflectedFace{}, \bisectionTree{}, \level{})$ where \simplex{}\ is the original simplex, \reflectedFace{}\ is a list of vertices, \bisectionTree{}\ is the bisection tree of the simplex \treeSimplex{}, and \level{}\ is the bisection level. The marked mesh is composed of tree-simplices. We create an empty marked mesh \markedMesh{}. For each simplex of the original mesh \mesh{}, we create a marked tree-simplex \treeSimplex{}, where \bisectionTree{}\ is the bisection tree of \simplex{}, \reflectedFace{}\ is an empty list, and the bisection level is 0. Then, we append the tree-simplex \treeSimplex{}\ to the marked mesh \markedMesh{}.

\subsection{First stage: tree-simplices}
\label{sec:firstStage}

In the first stage, we bisect the simplices using the bisection trees computed with the co-dimensional marking process. The first stage is used in the first $n-2$ bisection steps and, therefore, the generated simplices have at most descendant level $n-2$. Moreover, during the refinement process we store the new mid-vertices into \reflectedFace{}. Thus, in the second stage, we are able to map the generated simplices into a reflected mesh, a mesh type that it is also strongly compatible. When we refine adjacent simplices using their bisection trees, we will obtain a conformal mesh since, by construction, we enforce that a sub-simplex has the same bisection tree independently of the marked simplex it belongs to.

\begin{algorithm}[t!]
	\caption{Bisect a marked tree-simplex.}
	\label{alg:bisectStageOne}
	\begin{algorithmic}[1]
		\Require{\type{TreeSimplex} \treeSimplex{}}
		\Ensure{\type{TreeSimplex} \treeSimplex{1}, \type{TreeSimplex} \treeSimplex{2}}
		\Function{bisectStageOne}{\treeSimplex{}}
		\State $(\simplex{}, \reflectedFace{} , \bisectionTree{} , \level{}) = \treeSimplex{}$
		\State $\bisectionEdge{} = \Call{root}{\bisectionTree{}}$ \Comment{Bisection edge}
		\label{line:bisectionEdgeStageOne}
		\State $\simplex{1}, \reflectedFace{1}, \simplex{2}, \reflectedFace{2} = \Call{bisectTreeSimplex}{\simplex{}, \reflectedFace{}, \bisectionEdge{}, \level{}}$ 
		\label{line:bisectTreeSimplex1}
		\State \bisectionTree{1} = \Call{left}{\bisectionTree{}}; \bisectionTree{2} = \Call{right}{\bisectionTree{}} \Comment{Bisect tree}
		\label{line:bisectBalancedBisectionTree}
		\State $\level{1} = \level{} + 1$;  $\level{2} = \level{} + 1$ \Comment{Bisect level}
		\label{line:bisectLevelStageOne}
		\State $\treeSimplex{1} = (\simplex{1}, \reflectedFace{1}, \bisectionTree{1}, \level{1})$
		\label{line:buildMarkedTreeSimplex1}
		\State $\treeSimplex{2} = (\simplex{2}, \reflectedFace{2}, \bisectionTree{2}, \level{2})$
		\label{line:buildMarkedTreeSimplex2}
		\State \Return \treeSimplex{1}, \treeSimplex{2}
		\EndFunction
	\end{algorithmic}
\end{algorithm}

Algorithm \ref{alg:bisectStageOne} details the bisection process of a simplex in the first stage. First, we get the bisection edge, \bisectionEdge{}, taking the root of the bisection tree, \bisectionTree{}, see Line \ref{line:bisectionEdgeStageOne}. Then, we call the function that bisects a tree-simplex, see Algorithm \ref{alg:bisectTreeSimplex}. This function bisects \simplex{}\ into \simplex{1}\ and \simplex{2}, and returns two lists of vertices, \reflectedFace{1}\ and \reflectedFace{2}, see Line \ref{line:bisectTreeSimplex1}. Next, we proceed to bisect the bisection tree \bisectionTree{}\ generating two bisection trees \bisectionTree{1}\ and \bisectionTree{2}, see Line \ref{line:bisectBalancedBisectionTree}, that are the left and right branches of \bisectionTree{}, respectively. Recall that the branches have one level less than \bisectionTree{}.  We do the same with the level \level{}, and we obtain the levels \level{1}\ and \level{2}\ that are defined as $\level{} + 1$, see Line \ref{line:bisectLevelStageOne}. After that, we return the tree simplices \treeSimplex{1}\ and \treeSimplex{2}\ defined in Line \ref{line:buildMarkedTreeSimplex1} and Line \ref{line:buildMarkedTreeSimplex2}, respectively.

\begin{algorithm}[t!]
	\caption{Bisect a tree-simplex.}
	\label{alg:bisectTreeSimplex}
	\begin{algorithmic}[1]
		\Require{\type{Simplex} \simplex{}, \type{\level{}-List} \reflectedFace, \type{Edge} \bisectionEdge{}, \type{Level} \level{}}
		\Ensure{\type{Simplex} \simplex{1}, \type{$(\level{} + 1)$-List} \reflectedFace{1}, \type{Simplex} \simplex{2}, \type{$(\level{} + 1)$-List} \reflectedFace{2}}
		\Function{bisectTreeSimplex}{\simplex{}, \reflectedFace{}, \bisectionEdge{}, \level{}}
		\State $(\bvertex{0}, \bvertex{1}, \ldots, \bvertex{n})= \simplex{}$
		\label{line:copySimplex1}
		\State $(\multivertex{1,\level{}-1}{2,\level{}-1}, \ldots, \multivertex{1,0}{2,0}) = \reflectedFace{}$
		\label{line:copyListVertices}
		\State $(\vertex{1,\level{}},\vertex{2,\level{}}) = \bisectionEdge{}$
		\label{line:verticesBisectionEdge}
		\State $\multivertex{1,\level{}}{2,\level{}} = \Call{midVertex}{\vertex{1,\level{}},\vertex{2,\level{}}}$
		\label{line:makeMidVertex} 
		\State $(i_{1}, i_{2})$ = \Call{simplexVertices}{\simplex{}, \bisectionEdge{}}
		\label{line:simplexVertices}
		\State $\simplex{1}= (\bvertex{0}, \ldots, \bvertex{i_{2} - 1}, \multivertex{1,l}{2,l}, \bvertex{i_{2} + 1}, \ldots, \bvertex{n})$
		\label{line:child1}
		\State $\simplex{2}= (\bvertex{0}, \ldots, \bvertex{i_{1} - 1}, \multivertex{1,l}{2,l}, \bvertex{i_{1} + 1}, \ldots, \bvertex{n})$
		\label{line:child2}
		\State $\reflectedFace{1} = (\multivertex{1,\level{}}{2,\level{}},\multivertex{1,\level{}-1}{2,\level{}-1} \ldots, \multivertex{1,0}{2,0})$
		\label{line:reflected1}
		\State $\reflectedFace{2} = (\multivertex{1,\level{}}{2,\level{}},\multivertex{1,\level{}-1}{2,\level{}-1} \ldots, \multivertex{1,0}{2,0})$
		\label{line:reflected2}
		\State \Return \simplex{1}, \reflectedFace{1}, \simplex{2}, \reflectedFace{2}
		\EndFunction
	\end{algorithmic}
\end{algorithm}

To bisect a tree-simplex, we apply Algorithm \ref{alg:bisectTreeSimplex}. The inputs are a simplex, \simplex{}, an \level{}-list of vertices, \reflectedFace{}, a bisection edge, \bisectionEdge{}, and a descendant level, \level{}. We extract the vertex ids $(\bvertex{0}, \bvertex{1}, \ldots, \bvertex{n})$ of \simplex{}, see Line \ref{line:copySimplex1}. There are \level{}\ multi-ids of length two and $n+1-\level{}$ multi-ids of length one. This is so because at each bisection, we replace a multi-id of length one with a multi-id of length two, the latter multi-id corresponding to the mid-vertex.  In Line \ref{line:copyListVertices}, we extract the vertex ids of the vertices list.

Since the bisection edge belongs to the initial simplex, the length of the multi-ids of the vertices are one. Thus, we write $\edge{} = (\vertex{1,l},\vertex{2,l})$, see Line \ref{line:verticesBisectionEdge}, and we create the new mid-vertex \multivertex{1,l}{2,l}\ in Line \ref{line:makeMidVertex}. The new mid-vertex is identified using a multi-id of length two. Then, we obtain the local identifier of the vertices of \bisectionEdge{}\ inside the simplex \simplex{}, see Line \ref{line:simplexVertices}. After that, we define as children of \simplex{}\ the simplices \simplex{1}\ and \simplex{2}. In \simplex{1}\ we replace the $i_{2}$-th local vertex by the new vertex \multivertex{1,l}{2,l}, see Line \ref{line:child1}. We proceed in the same manner for \simplex{2}\ substituting the $i_{1}$-th local vertex by the new vertex, see Line \ref{line:child2}.  Next, we add the new vertex \multivertex{1,l}{2,l}\ at the beginning of the lists \reflectedFace{1}\ and \reflectedFace{2}, see Lines \ref{line:reflected1} and \ref{line:reflected2}, respectively. 

Note that \reflectedFace{1}\ and \reflectedFace{2}\ are the same list on the current bisection step. Nevertheless, in the next bisection step, we may append a different vertex to each list. Moreover, a tree-simplex with descendant level \level{}\ has a sorted list composed of \level{}\ vertices. 
Since a tree-simplex descendant level is at most $n-1$, the list of vertices contains at most $n-1$ vertices.

\subsection{Second stage: casting to Maubach}
\label{sec:secondStage}

\begin{algorithm}[t!]
	\caption{Bisect to Maubach}
	\label{alg:bisectToMaubach}
	\begin{algorithmic}[1]
		\Require{\type{TreeSimplex} \treeSimplex{}}
		\Ensure{\type{MaubachSimplex} \maubachSimplex{1}, \type{MaubachSimplex} \maubachSimplex{2}}
		\Function{bisectToMaubach}{\treeSimplex{}}
		\State $(\simplex{}, \reflectedSimplex{}, \bisectionTree{}, \level{}) = \treeSimplex{}$
		\State $\bisectionEdge{} = \Call{root}{\bisectionTree{}}$ 
		\label{line:bisectionEdgeCastToMaubach}
		\State $\simplex{1},\reflectedFace{1},\simplex{2},\reflectedFace{2} = \Call{bisectTreeSimplex}{\simplex{},\reflectedFace{},\bisectionEdge{},\level{}}$ 
		\label{line:bisectTreeSimplexCastToMaubach}
		\State $\reflectedSimplex{1}, \reflectedSimplex{2} = \Call{castToMaubach}{\bisectionEdge{},\reflectedFace{1},\reflectedFace{2}}$ 
		\label{line:castToMaubach}
		\State $\tagMaubach{1} = n$; $\tagMaubach{2} = n$
		\label{line:setTagCastToMaubach}
		\State $\level{1} = \level{} + 1$; $\level{2} = \level{} + 1$ 
		\label{line:bisectLevelCastToMaubach}
		\State $\maubachSimplex{1} = (\reflectedSimplex{1},\tagMaubach{1}, \level{1})$
		\State $\maubachSimplex{2} = (\reflectedSimplex{2}, \tagMaubach{2}, \level{2})$
		\State \Return $\maubachSimplex{1}, \maubachSimplex{2}$
		\EndFunction
	\end{algorithmic}
\end{algorithm}

We next introduce the second stage of our bisection method for simplices. In this stage, after bisecting a simplex, we reorder the vertices of the bisected simplices to obtain a reflected mesh. Thus, in the third stage, we can apply Maubach's algorithm to further refine the mesh.

The second stage is used when the descendant level of a tree-simplex, \treeSimplex{}, is $\level{} = n-1$. At this step, \reflectedFace{}\ is a $(n-1)$-list of  vertices that have been accumulated in the previous $(n-1)$ steps of the first stage. Finally, \bisectionTree{}\ is a bisection tree that only contains a single vertex. That is, \bisectionTree{}\ is a leaf where the consistent bisection edge $\bisectionEdge{} = \edgeVertices{1,n-1}{2,n-1}$ is the root.

In the second stage of our proposed bisection method, the input is a tree-simplex $\treeSimplex{} = (\simplex{}, \reflectedFace{}, \bisectionTree{}, \level{})$, and the outputs are two Maubach simplices \maubachSimplex{1}\ and \maubachSimplex{2}. A Maubach simplex is a 3-tuple $\maubachSimplex{} = (\reflectedSimplex{}, \tagMaubach{}, \level{})$, where \reflectedSimplex{}\ is an equivalent simplex, but it is reordered to properly contribute to a reflected mesh, \tagMaubach{}\ is a Maubach's integer tag, and \level{}\ is the descendant level. 

\begin{algorithm}[t!]
	\caption{Cast to Maubach.}
	\label{alg:castToMaubach}
	\begin{algorithmic}[1]
		\Require{\type{Edge} \bisectionEdge{}, $n$\type{-List} \reflectedFace{1}, $n$\type{-List} \reflectedFace{2}}
		\Ensure{$n$\type{-Simplex} \reflectedSimplex{1}, $n$\type{-Simplex} \reflectedSimplex{2}}
		\Function{castToMaubach}{\bisectionEdge{}, \reflectedFace{1}, \reflectedFace{2}}
		\State $\edgeVertices{1,n-1}{2,n-1} = \bisectionEdge{}$
		\State $(\multivertex{1,n-1}{2,n-1}, \ldots, \multivertex{1,0}{2,0}) = \reflectedFace{1}$
		\State $(\multivertex{1,n-1}{2,n-1}, \ldots, \multivertex{1,0}{2,0}) = \reflectedFace{2}$
		\State $\reflectedSimplex{1} = (\vertex{1,n-1}, \multivertex{1,n-1}{2,n-1} , \ldots, \multivertex{1,0}{2,0})$
		\State $\reflectedSimplex{2} = (\vertex{2,n-1}, \multivertex{1,n-1}{2,n-1} , \ldots , \multivertex{1,0}{2,0})$
		\State \Return \reflectedSimplex{1}, \reflectedSimplex{2}
		\EndFunction
	\end{algorithmic}
\end{algorithm}
First, we obtain the consistent bisection edge, $\bisectionEdge{} = \edgeVertices{1,n-1}{2,n-1}$, Line \ref{line:bisectionEdgeCastToMaubach}. Then, we bisect the simplex \simplex{}\ using the function that bisects tree-simplices, generating two simplices \simplex{1}\ and \simplex{2}\ and the sorted $n$-lists of vertices \reflectedFace{1}\ and \reflectedFace{2}. After that, we reorder the simplices to be able to apply Maubach's algorithm, see Line \ref{line:castToMaubach}. After generating the two simplices \reflectedSimplex{1}\ and \reflectedSimplex{2}\ that define a reflected neighbors configuration, we bisect the descendant level \level{}\ generating two descendant levels $\level{1} = n$ and $\level{2} = n$, see Line \ref{line:bisectLevelCastToMaubach}.  Then, we set the Maubach tags $\tagMaubach{1} = n$ and $\tagMaubach{2} = n$, see Line \ref{line:setTagCastToMaubach}. Finally, we create the Maubach's simplices $\maubachSimplex{1} = (\reflectedSimplex{1}, \tagMaubach{1}, \level{1})$ and $\maubachSimplex{2} = (\reflectedSimplex{2}, \tagMaubach{2}, \level{2})$.

To cast the simplices to obtain a reflected configuration, we apply Algorithm \ref{alg:castToMaubach}. The inputs are the bisection edge, and two lists of vertices. The outputs are two reordered simplices in a reflected configuration. This algorithm adds the vertices of the bisection edge \vertex{1,n-1}\ and \vertex{2,n-1}\ to \reflectedFace{1}\ and \reflectedFace{2}, respectively, generating two sorted $(n+1)$-list. The lists \reflectedFace{1}\ and \reflectedFace{2} contain the same vertices that \simplex{1}\ and \simplex{2}, respectively, but in a different order. The order of vertices induced by \reflectedFace{1}\ and \reflectedFace{2} leads to a reflected mesh. 
Next, we create the simplices \reflectedSimplex{1}\ and \reflectedSimplex{2} by casting the $(n+1)$-sorted list of vertices \reflectedFace{1}\ and \reflectedFace{2} into simplices, respectively.

\subsection{Third stage: Maubach's bisection}
\label{sec:thirdStage}

\begin{algorithm}[t!]
	\caption{Adapted Maubach's algorithm.}
	\label{alg:bisectMaubach}
	\begin{algorithmic}[1]
		\Require{\type{MaubachSimplex} \maubachSimplex{}}
		\Ensure{\type{MaubachSimplex} \maubachSimplex{1}, \type{MaubachSimplex} \maubachSimplex{2}}
		
		\Function{bisectMaubach}{\maubachSimplex{}}
		\State $((\bvertex{0}, \bvertex{1}, \ldots, \bvertex{n}), \tagMaubach{}, \level{}) = \maubachSimplex{}$
		\State $\bwertex{} = \Call{midVertex}{\bvertex{0}, \bvertex{\tagMaubach{}}}$
		\label{line:midVertexMaubach}
		\State $\reflectedSimplex{1} = (\bvertex{0}, \ldots, \bvertex{d-1}, \bwertex{}, \bvertex{d+1}, \ldots, \bvertex{n})$
		\label{line:bisectMaubachSimplex1}
		\State $\reflectedSimplex{2} = (\bvertex{1}, \ldots, \bvertex{d}, \bwertex{}, \bvertex{d+1}, \ldots, \bvertex{n})$
		\label{line:bisectMaubachSimplex2}
		\State Set 
		$
		\tagMaubach{}^{\prime} = \left\{
		\begin{array}{ll} 
		\tagMaubach{}-1, & \tagMaubach{} > 1 \\ 
		n, & \tagMaubach{} = 1 
		\end{array}
		\right.
		$
		\label{line:bisectTagMaubach}
		\State $\tagMaubach{1} = \tagMaubach{}^{\prime}$; $\tagMaubach{2} = \tagMaubach{}^{\prime}$
		\label{line:bisectTagsMaubach}
		\State $\level{1} = \level{} + 1$; $\level{2} = \level{} + 1$
		\label{line:bisectLevelMaubach}
		\State $\maubachSimplex{1} = (\reflectedSimplex{1}, \tagMaubach{1}, \level{1})$
		\label{line:buildMarkedMaubachSimplex1}
		\State $\maubachSimplex{2}= (\reflectedSimplex{2}, \tagMaubach{2}, \level{2})$
		\label{line:buildMarkedMaubachSimplex2}
		\State \Return $\maubachSimplex{1}, \maubachSimplex{2}$
		\EndFunction
	\end{algorithmic}
\end{algorithm}

Following, we describe the third stage of our bisection algorithm. In this stage, we use Maubach's algorithm to favor the conformity, finiteness, stability, and locality properties.

We reinterpret Maubach's algorithm using tagged simplices and multi-ids in Algorithm \ref{alg:bisectMaubach}. The input is a Maubach simplex, $\maubachSimplex{} = (\reflectedSimplex{}, \tagMaubach{}, \level{})$, and the outputs are two Maubach simplices, \maubachSimplex{1}\ and \maubachSimplex{2}. First, we generate the new vertex \bwertex\ as the mid-vertex of \bvertex{0}\ and \bvertex{d}, see Line \ref{line:midVertexMaubach}. That is, the bisection edge is $\bisectionEdge{} = (\bvertex{0},\bvertex{d})$. Then, we bisect \reflectedSimplex{}\ and generate the children \reflectedSimplex{1}\ and \reflectedSimplex{2}, see Lines \ref{line:bisectMaubachSimplex1} and \ref{line:bisectMaubachSimplex2}, respectively. After that, we set the new tag $\tagMaubach{}^{\prime}$ for the children simplices, see Line \ref{line:bisectTagMaubach}. Thus, we define the tags \tagMaubach{1}\ and \tagMaubach{2}\ as $\tagMaubach{}^{\prime}$, see Line \ref{line:bisectTagsMaubach}. Analogously, we bisect the levels \level{1}\ and \level{2}, see Line \ref{line:bisectLevelMaubach}. Finally, we create two Maubach's simplices \maubachSimplex{1}\ and \maubachSimplex{2}, see Lines \ref{line:buildMarkedMaubachSimplex1} and \ref{line:buildMarkedMaubachSimplex2}, respectively.

In this algorithm, we use the notation $\simplex{} = (\bvertex{0}, \bvertex{1}, \ldots, \bvertex{n})$ because any of the multi-ids can be of length two or higher. This is so since if we call local refine after previous local refinements, our marked mesh might contain Maubach simplices with tags in the range $1 \leq d \leq n - 2$. For this tag range, Maubach's algorithm bisects edges of newer generations before bisecting all the original edges of the Maubach simplex \cite{alkamper2018weak}. Therefore, we can have multi-ids of length higher than two. 

\section{Examples}
\label{sec:examples}

We present several examples to illustrate that our proposed algorithm bisects unstructured simplicial meshes, locally adapts conformal meshes, generates a finite number of similarity classes, and leads to lower-bounded quality meshes. For all the examples, we have computed the shape quality of the mesh.
To calculate the shape quality for $n$-dimensional simplices, we use the expression
\[
\dfrac{\tr (S^{t} S)}{n \det (S)^{2/n}},
\]
where $S$ is the Jacobian of the affine mapping between the ideal equilateral simplex and the physical simplex \cite{knupp2001algebraic, liu1994shape}.
We plot the minimum and maximum shape quality of the mesh in each refinement step to illustrate that the minimum quality is lower bounded and cycles. 

To verify the results, we need the capabilities to check the conformity and the reflectivity of some of the meshes generated during the refinement process. To check the conformity, we check that all the interior faces of \mesh{k}\ are shared only by two simplices and that all the boundary faces of mesh \mesh{k}\ are descendants of the original boundary faces. See more details in \ref{sec:conformity}. To check the reflectivity of \mesh{k}, we check that all the neighboring simplices are reflected neighbors through the shared face. See more details in \ref{sec:reflectivity}. Using these verification capabilities, we can check that for any refinement set, including random sets, the method leads to conformal meshes. Specifically, the bisection method refines the marked simplices, and it refines the required surrounding simplices to maintain conformity.

All the results have been obtained on a MacBook Pro with one dual-core Intel Core i5 CPU, with a clock frequency of 2.7GHz, and a total memory of 16GBytes. As a proof of concept, a mesh refiner has been fully developed in Julia 1.4. The Julia prototype code is sequential (one execution thread), corresponding to the implementation of the method presented in this work. All the unstructured initial meshes are generated with the \texttt{distmesh} \cite{persson2004simple} algorithm, and all the structured initial meshes are generated with the Coxeter-Freudenthal-Kuhn \cite{coxeter1934discrete, freudenthal1942simplizialzerlegungen, kuhn1960some} algorithm. We recall that after each refinement to conformity, we perform a renumber of the vertices of the mesh \mesh{k}\ in order to have only multi-ids of length one, see \ref{sec:renumber}.

\subsection{First two stages on unstructured meshes: conformity and reflectivity}

The goal of this example is to propose a methodology to check the correctness of the algorithm implementation. Thus, let $\mesh{0}^{n}$ be a conformal unstructured $n$-simplicial mesh and consider $n$ uniform refinements using our proposed bisection algorithm.  That is, we apply our bisection method until the second stage is executed. At this point of the algorithm, we have reordered the vertices of the simplices and we should obtain a conformal and reflected mesh. We check if the generated $n$-dimensional mesh $\mesh{n}^{n}$ is conformal and reflected, conditions that are sufficient to apply Maubach's algorithm. The proposed algorithms to check the conformity and reflectivity of $\mesh{n}^{n}$ are depicted in \ref{sec:conformity} and \ref{sec:reflectivity}, respectively. 

\begin{figure}[t!]
	\centering
	\begin{tabular}{ccc}
		\subfigure[]{
			\includegraphics[width=0.27\textwidth]{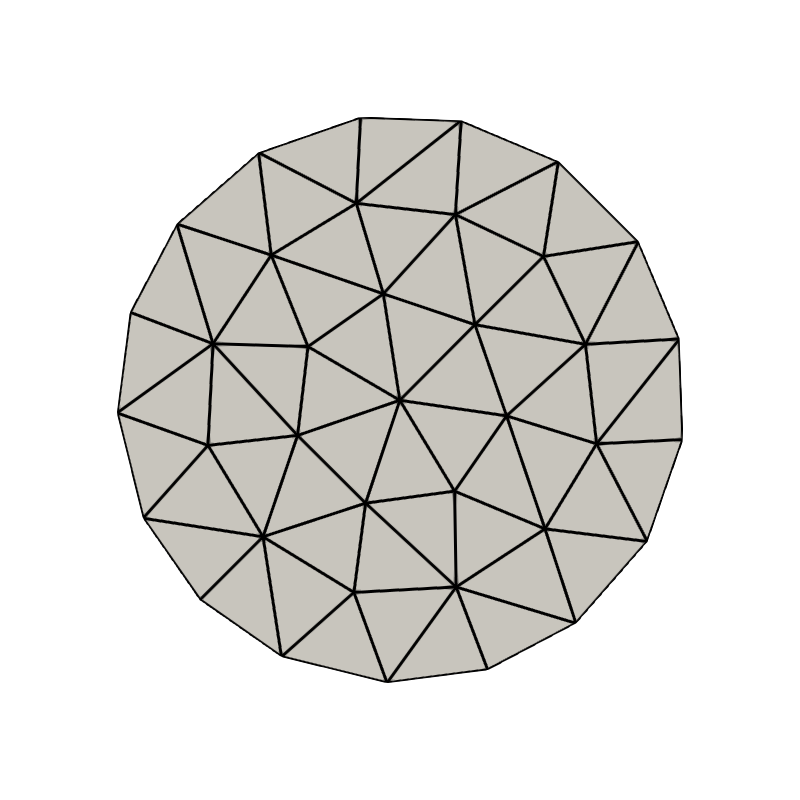}
			\label{fig:hypersphere2D_0}
		} 
		&
		\subfigure[]{
			\includegraphics[width=0.27\textwidth]{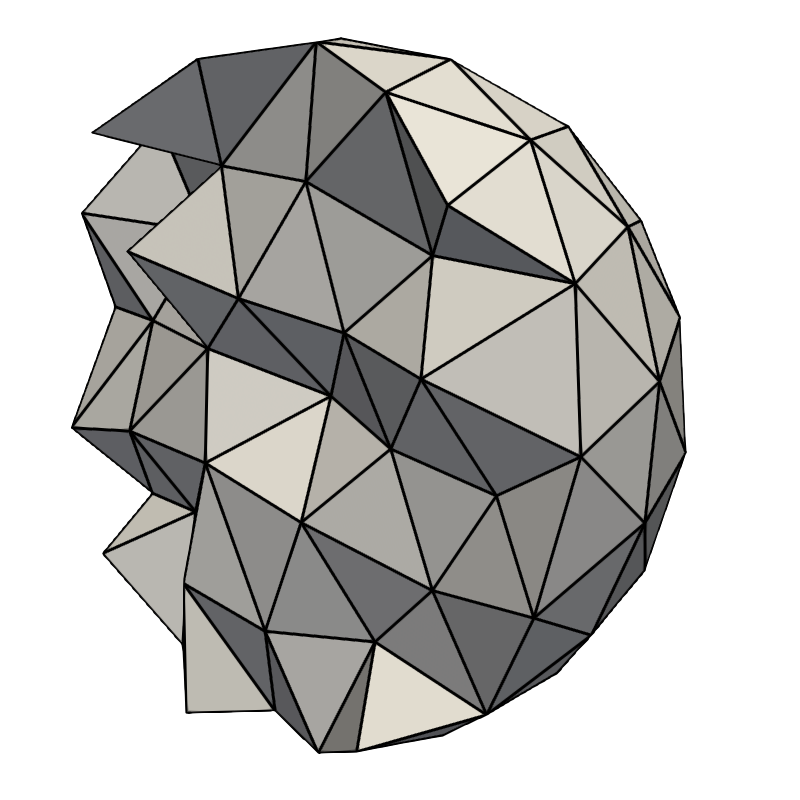}
			\label{fig:hypersphere3D_0}
		}
		&
		\subfigure[]{
			\includegraphics[width=0.27\textwidth]{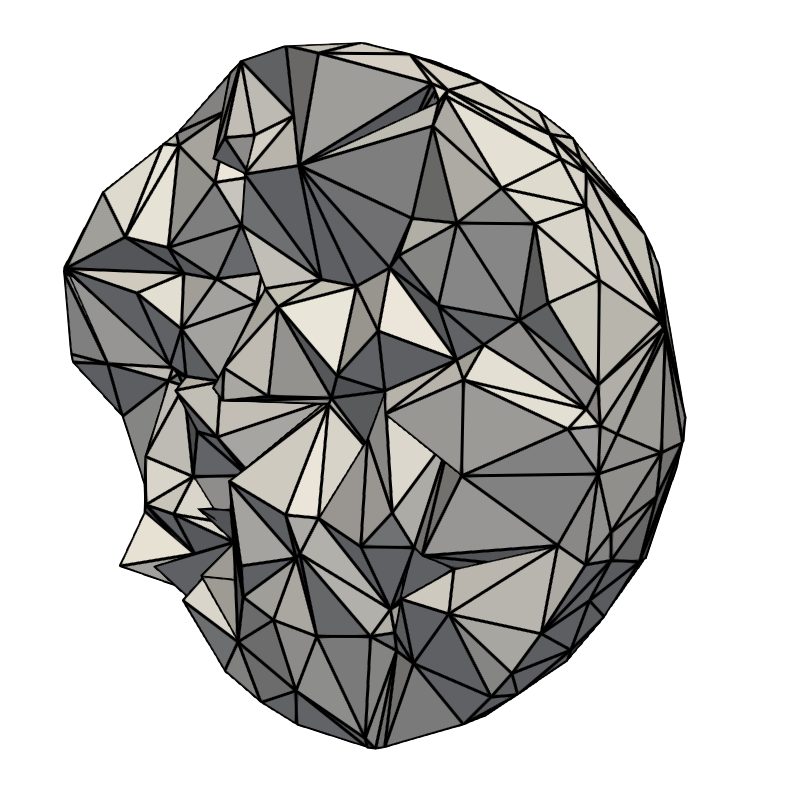}
			\label{fig:hypersphere4D_0}
		} 
		\\
		\subfigure[]{
			\includegraphics[width=0.27\textwidth]{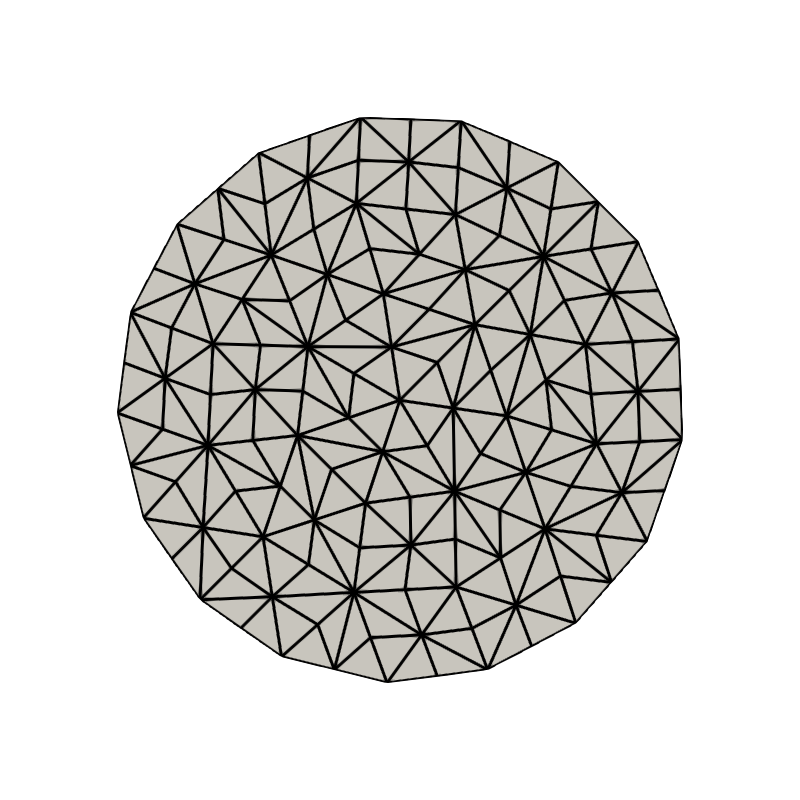}
			\label{fig:hypersphere2D_1}
		} 
		&
		\subfigure[]{
			\includegraphics[width=0.27\textwidth]{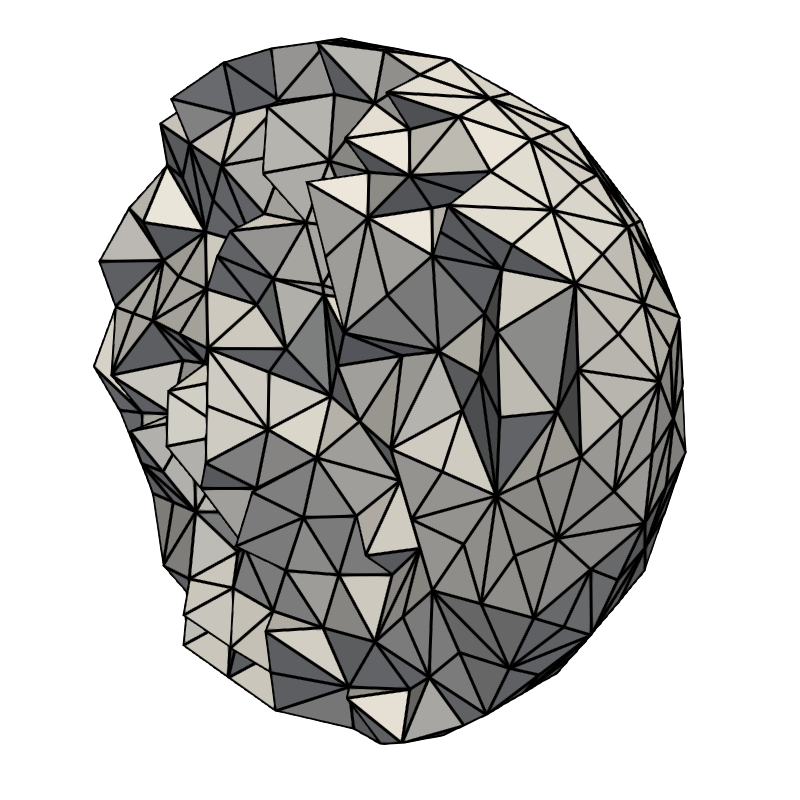}
			\label{fig:hypersphere3D_1}
		}
		&
		\subfigure[]{
			\includegraphics[width=0.27\textwidth]{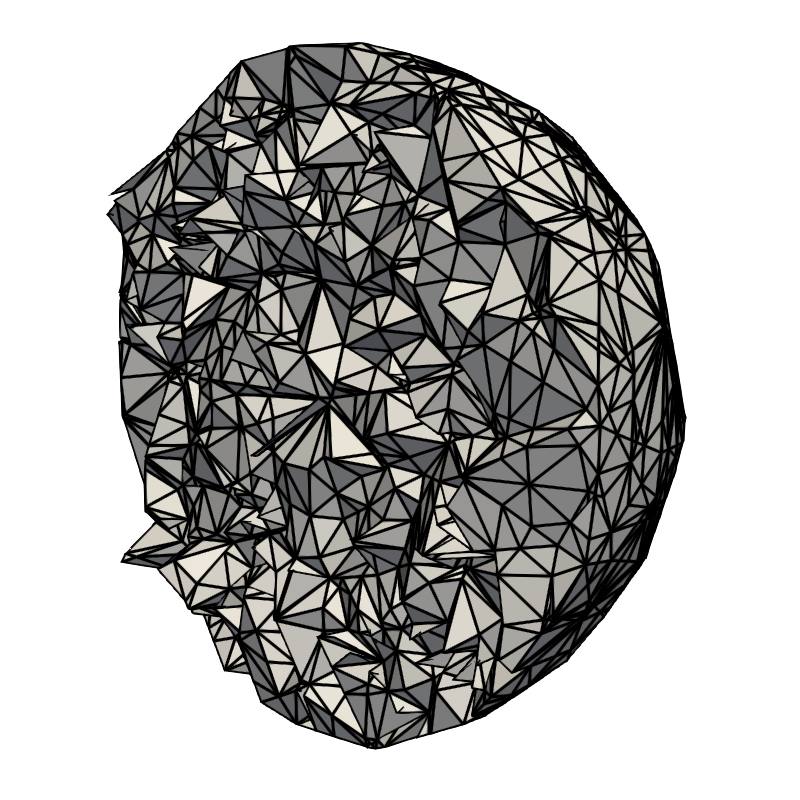}
			\label{fig:hypersphere4D_1}
		} 
	\end{tabular}
	\caption{Initial conformal $n$-dimensional meshes and conformal reflected $n$-dimensional meshes after the first two bisection stages. \subref{fig:hypersphere2D_0} Initial conformal triangular mesh $\mesh{0}^{2}$ and \subref{fig:hypersphere2D_1} conformal reflected triangular mesh $\mesh{2}^{2}$ after two uniform refinements. \subref{fig:hypersphere3D_0} Initial conformal tetrahedral mesh $\mesh{0}^{3}$ and \subref{fig:hypersphere3D_1} conformal reflected tetrahedral mesh $\mesh{3}^{3}$ after three uniform refinements. Volume slice with the hyper-plane $t = 0$ of \subref{fig:hypersphere3D_0} the initial conformal pentatopic mesh $\mesh{0}^{4}$ and \subref{fig:hypersphere4D_1} the conformal reflected pentatopic mesh $\mesh{4}^{4}$ after four uniform refinements.}
	\label{fig:conformalNUniformRefinements}
\end{figure} 
To verify our proposed algorithm, we present an unstructured example in different dimensions that allows us to check that the obtained results are the expected ones.  Let \closedBall{n}\ be the $n$-dimensional closed ball of radius $1$ centered at the origin, defined as the points such that $\norm{\vec{x}} \leq 1$. We approximate the domains \closedBall{2}, \closedBall{3}, \closedBall{4}, and \closedBall{5} with the simplicial meshes $\mesh{0}^{2}$, $\mesh{0}^{3}$, $\mesh{0}^{4}$, and $\mesh{0}^{5}$, respectively. To obtain equivalent mesh resolution, we set the edge length to 0.3 in all the cases. The initial meshes are shown in the first row, see Figure \ref{fig:conformalNUniformRefinements}.

After generating $\mesh{0}^{n}$ for each dimension, we apply $n$ uniform refinements using our bisection algorithm, and we obtain the meshes $\mesh{n}^{n}$ for $n = 2, 3, 4$, and $5$. The meshes $\mesh{n}^{n}$ have $2^{n}N^{n_{e}}$ simplices, where $N^{n_{e}}$ is the number of simplices of the initial mesh $\mesh{0}^{n}$. In Figures \ref{fig:hypersphere2D_1} and \ref{fig:hypersphere3D_1}, we show the meshes obtained for the two- and three- dimensional cases, respectively. Moreover, in Figure \ref{fig:hypersphere4D_1}, we show a 3-dimensional slice of the 4-dimensional mesh.

After the second stage, all the meshes are conformal and reflected, and thus they are strongly compatible. For this reason, we can apply the third stage, which is Maubach's algorithm, to further refine the meshes.

\subsection{Uniform bisection: minimum quality is lower bounded and cycles}
\label{example:minimumQuality}

\begin{figure}[t!]
	\centering
	\begin{tabular}{cc}
		\subfigure[]{
			\includegraphics[width=0.44\textwidth]{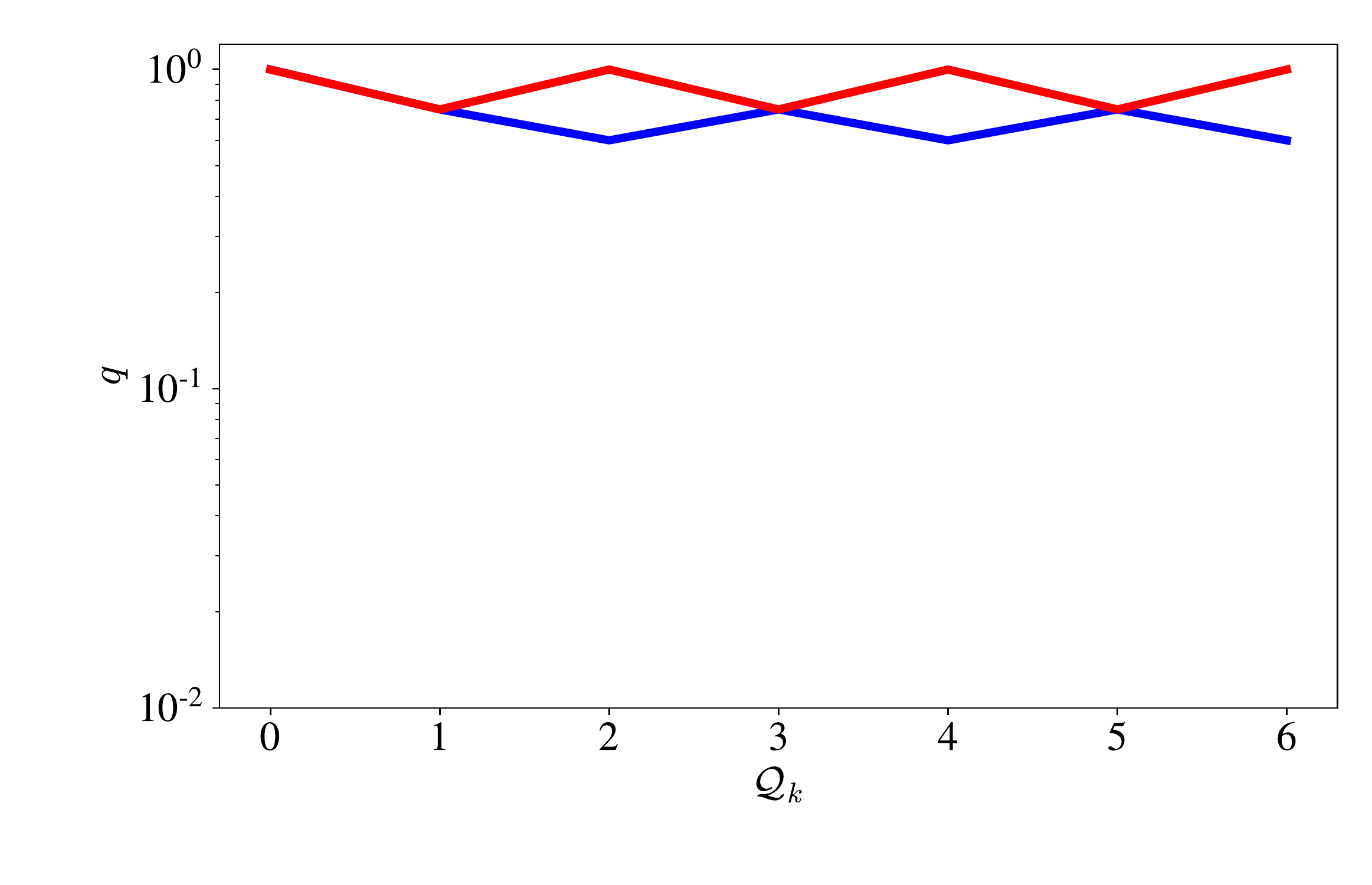}
			\label{fig:equilateral2D}
		} 
		&
		\subfigure[]{
			\includegraphics[width=0.44\textwidth]{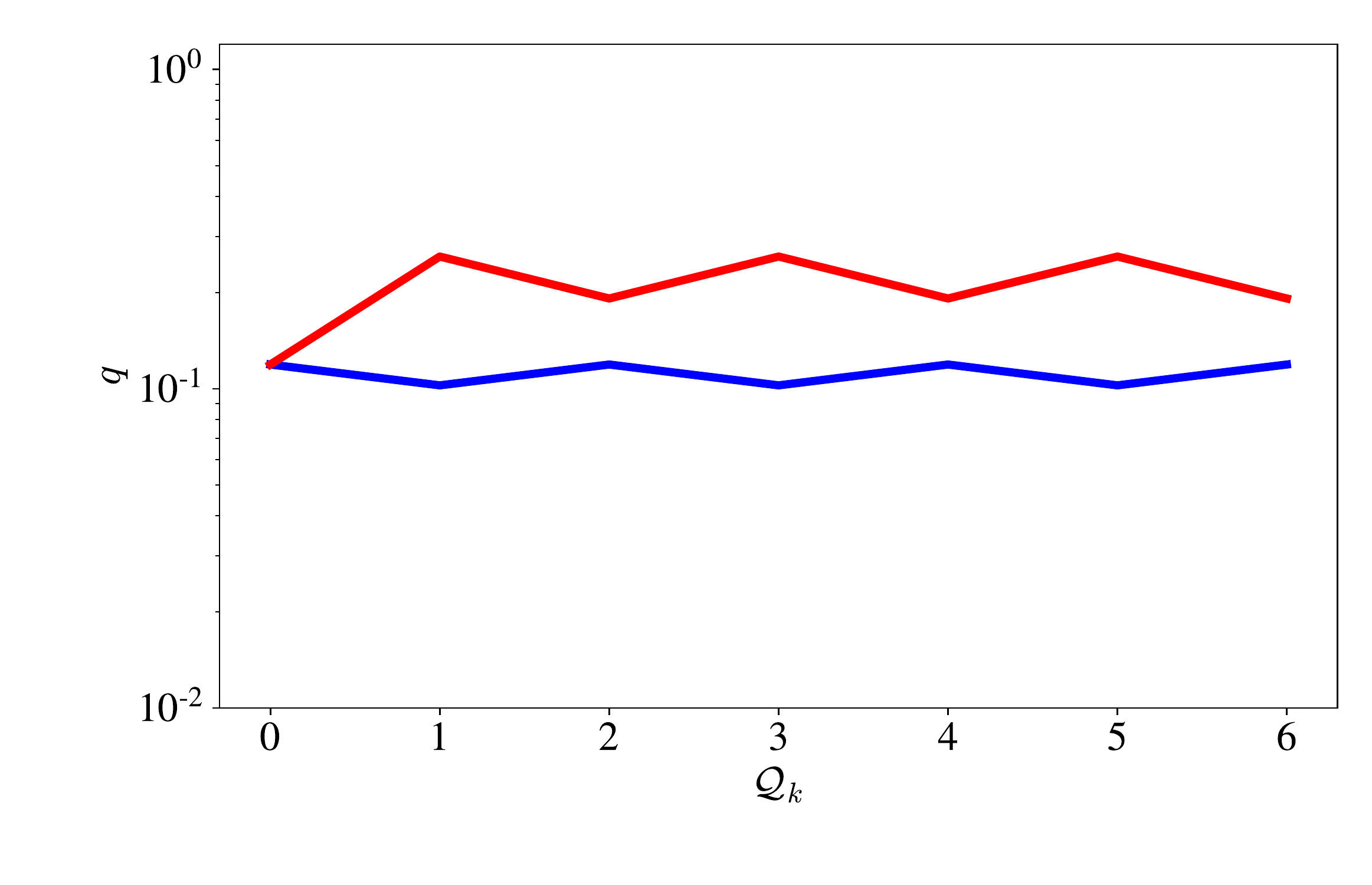}
			\label{fig:random2D}
		}
		\\
		\subfigure[]{
			\includegraphics[width=0.44\textwidth]{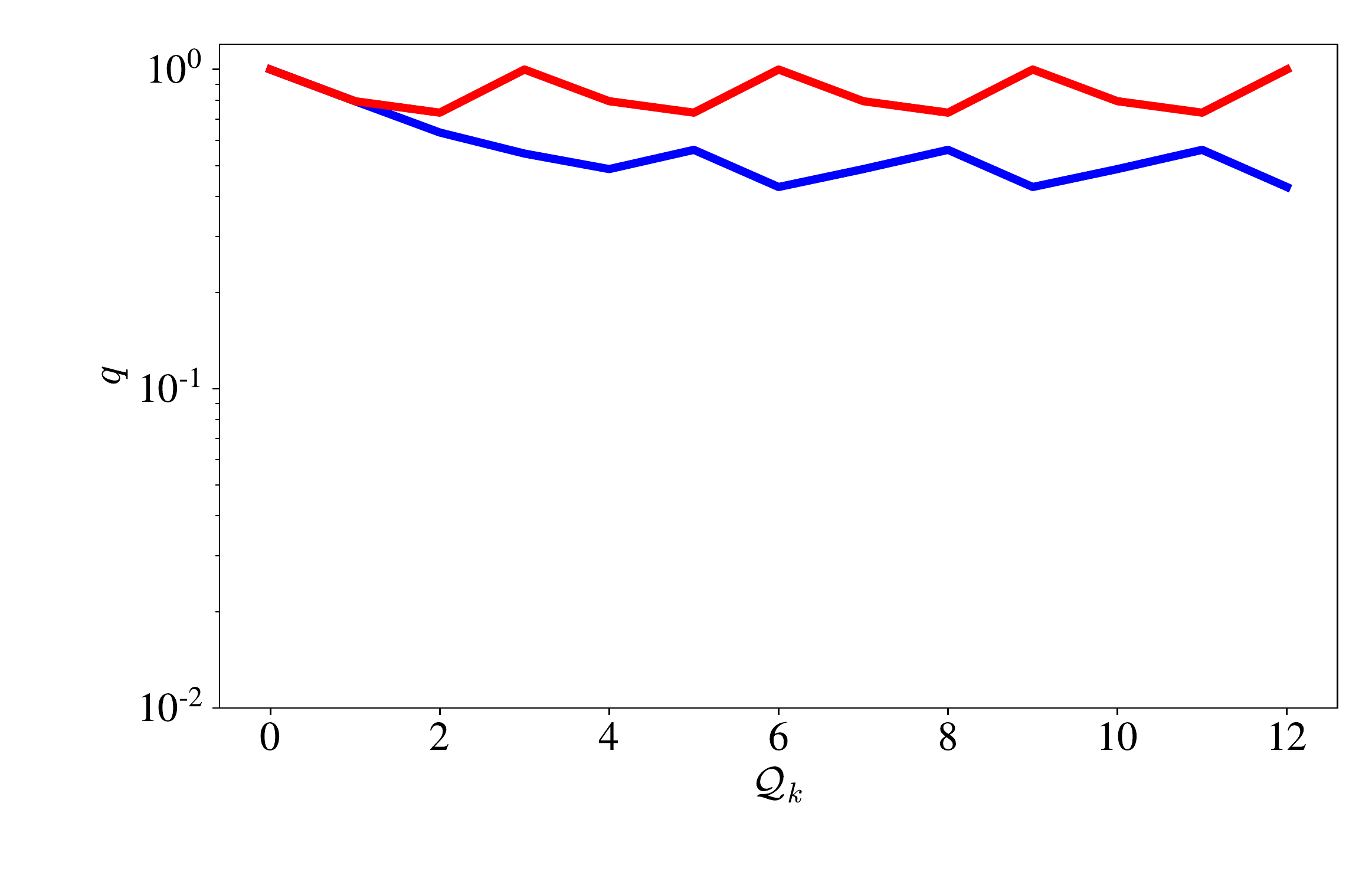}
			\label{fig:equilateral3D}
		}
		&
		\subfigure[]{
			\includegraphics[width=0.44\textwidth]{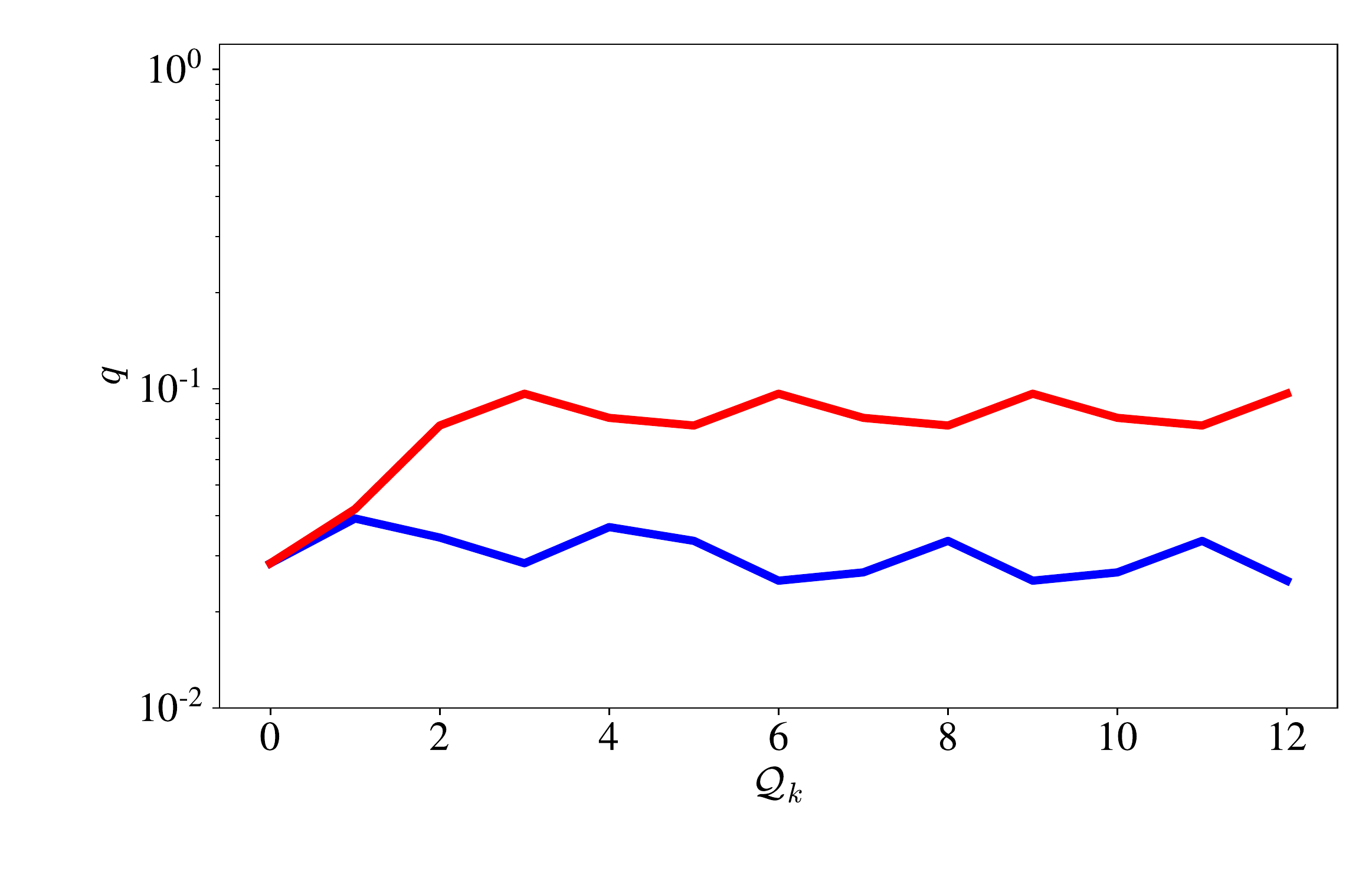}
			\label{fig:random3D}
		}
		\\ 
		\subfigure[]{
			\includegraphics[width=0.44\textwidth]{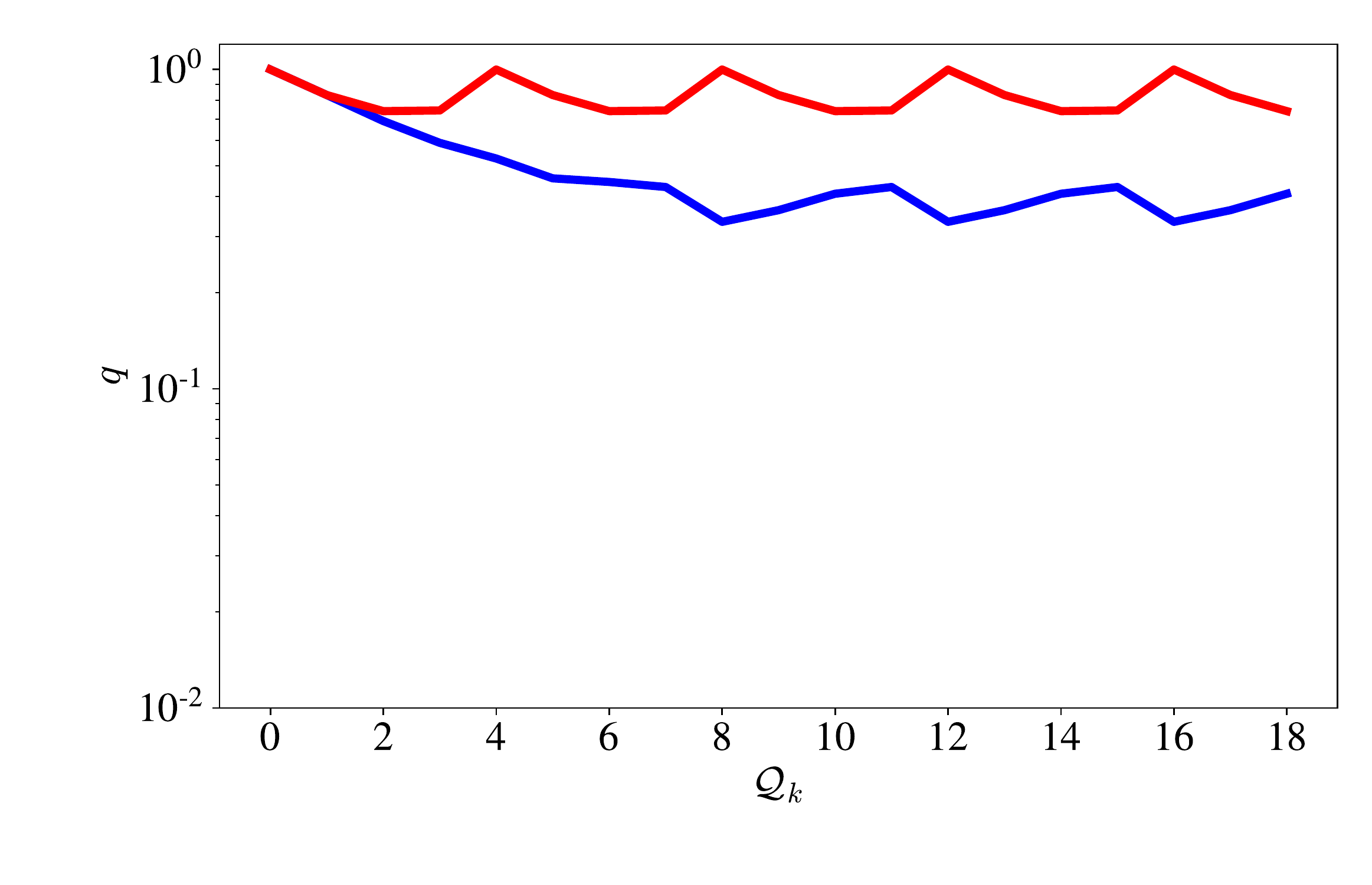}
			\label{fig:equilateral4D}
		} 
		&
		\subfigure[]{
			\includegraphics[width=0.44\textwidth]{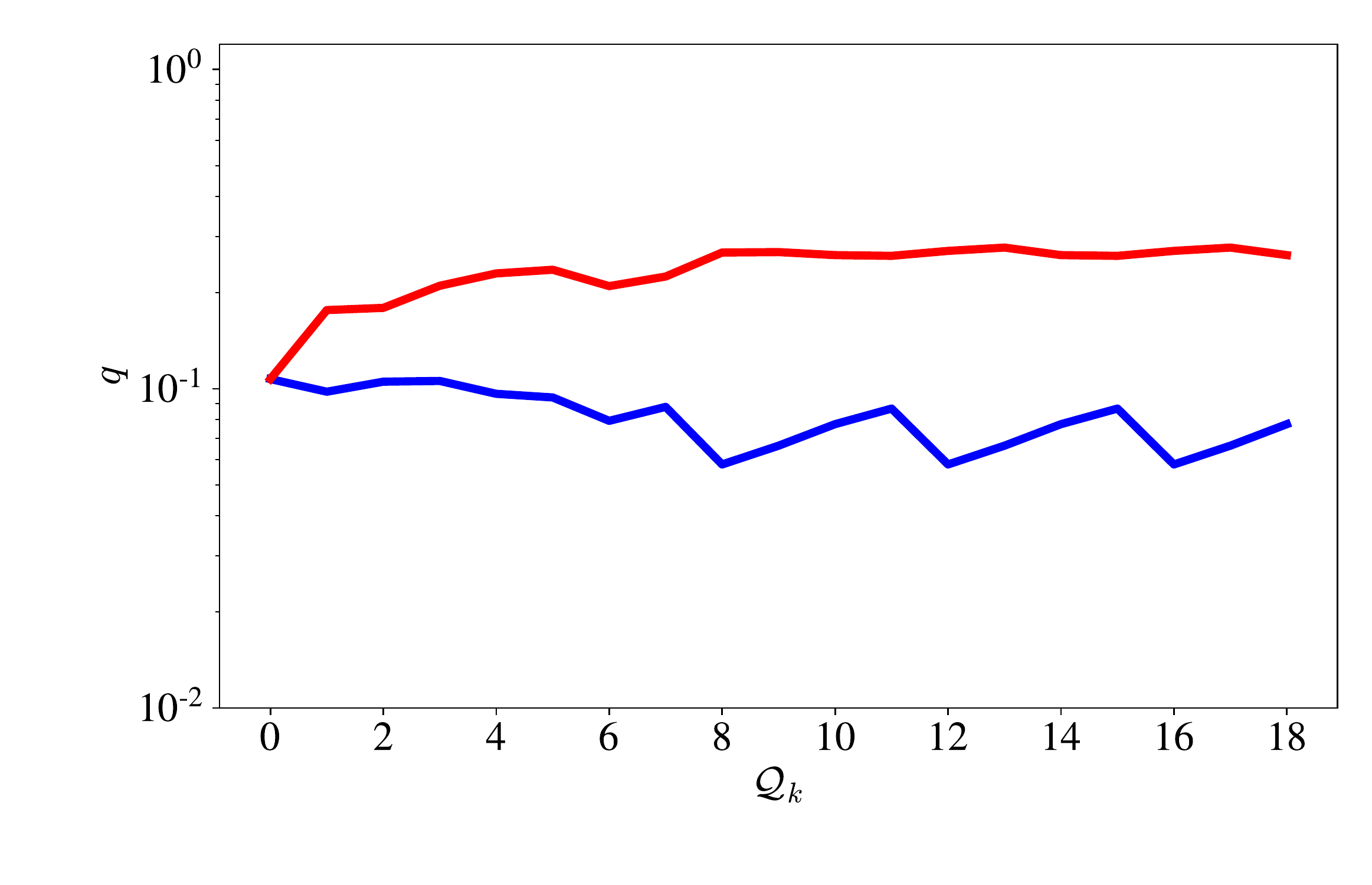}
			\label{fig:random4D}
		}
	\end{tabular}
	\caption{Minimum (blue) and maximum (red) quality cycles for uniform refinements. 
		In columns, initial simplex: \subref{fig:equilateral2D}, \subref{fig:equilateral3D} and \subref{fig:equilateral4D} equilateral, \subref{fig:random2D}, \subref{fig:random3D} and \subref{fig:random4D} irregular simplex.
		In rows, simplex dimension: \subref{fig:equilateral2D} and \subref{fig:random2D} triangles; 
		\subref{fig:equilateral3D} and \subref{fig:random3D} tetrahedra;
		\subref{fig:equilateral4D} and \subref{fig:random4D} pentatopes.}
	\label{fig:exampleQualityCycles}
\end{figure}

Following, we show that the minimum quality is lower bounded and cycles. To this end, we uniformly refine a single simplex several times. To illustrate the quality cycles, we perform a series of refinements to ensure that the method completes two additional cycles of Maubach's method.

Figure \ref{fig:exampleQualityCycles} plots the evolution of the minimum (blue line) and maximum (red line) qualities during the uniform refinement process. Each column of Figure \ref{fig:exampleQualityCycles} corresponds to a dimensional case, starting from 2D and ending in 4D. Analogously, the first and second rows of Figure \ref{fig:exampleQualityCycles} correspond to an equilateral and an irregular simplex, respectively. Note that in all cases, the minimum quality is lower-bounded, and the maximum and minimum qualities cycle with a period of $n$ steps.

For the 2-dimensional case, we illustrate in Figures \ref{fig:equilateral2D} and \ref{fig:random2D} the evolution of the minimum and maximum qualities of the meshes obtained by uniformly bisecting an equilateral triangle and an irregular triangle, respectively. For this case, we perform six uniform refinements. Since our method marks a triangle as a tagged triangle with $d = 1$ or $d=2$, it is analogous to Maubach's algorithm and, with two uniform refinements, all the similarity classes of a triangle are generated. At the second iteration, both the minimum and the maximum qualities start to cycle with a period of two steps.

For the 3-dimensional case, we show in Figures \ref{fig:equilateral3D} and \ref{fig:random3D} the evolution of the minimum and maximum qualities of the meshes obtained by uniformly bisecting an equilateral tetrahedron and an irregular tetrahedron, respectively. We perform twelve uniform refinements to generate all the similarity classes. For the equilateral tetrahedron, the marking process generates a bisection tree that is equivalent to a tagged tetrahedron with tag $d=2$ or, equivalently, a planar tetrahedron $P_{u}$, as denoted in \cite{arnold2000locally}. At iteration six, the process achieves the minimum quality, and both the minimum and maximum qualities cycle with a period of three steps, see Figure \ref{fig:equilateral3D}. We obtain similar results for the irregular tetrahedron, see Figure \ref{fig:random3D}. Specifically, the minimum quality is achieved at iteration six, and both the minimum and maximum qualities cycle with a period of three steps.

For the 4-dimensional case, we show in Figures \ref{fig:equilateral4D} and \ref{fig:random4D} the evolution of the minimum and maximum qualities of the meshes obtained by uniformly bisecting an equilateral pentatope and an irregular pentatope, respectively. We performed eighteen uniform refinements. In the case of the equilateral pentatope, the minimum quality is achieved at iteration eight, and both the minimum and maximum quality cycle with a period of four steps, see Figure \ref{fig:equilateral4D}. When bisecting the irregular pentatope, the minimum quality is obtained at iteration twelve, and both the minimum and maximum qualities cycle with a period of four steps, see Figure \ref{fig:random4D}.

In all the cases we perform enough uniform bisection steps to generate all the similarity classes and show some full refinement cycles of length $n$. Moreover, we reach the minimum and maximum mesh quality in a finite number of steps, qualities that are repeated every $n$ steps. This illustrates that the mesh quality does not degenerate relative to the initial quality under successive refinement, and thus, the method is stable.

\subsection{Local refine of a 4D structured mesh: equivalency to Maubach's method}

The main goal of this example is to illustrate that our algorithm is equivalent to newest vertex bisection when we refine a structured mesh. To this end, we recreate the first example from Maubach \cite{maubach1995local} and Arnold \etal\ \cite{arnold2000locally} but we extend it to four dimensions. Let $[0,1]^{4}$ be the unit hypercube and consider its subdivision into 16 sub-hypercubes. We subdivide into 24 pentatopes each sub-hypercube using Coxeter-Freudenthal-Kuhn \cite{coxeter1934discrete, freudenthal1942simplizialzerlegungen, kuhn1960some} algorithm, generating a 4D pentatopic mesh, \mesh{0}, composed of 384 pentatopes and 81 vertices.

\begin{figure}[t!]
	\centering
	\begin{tabular}{ccc}
		\subfigure[]{
			\includegraphics[width=0.275\textwidth]{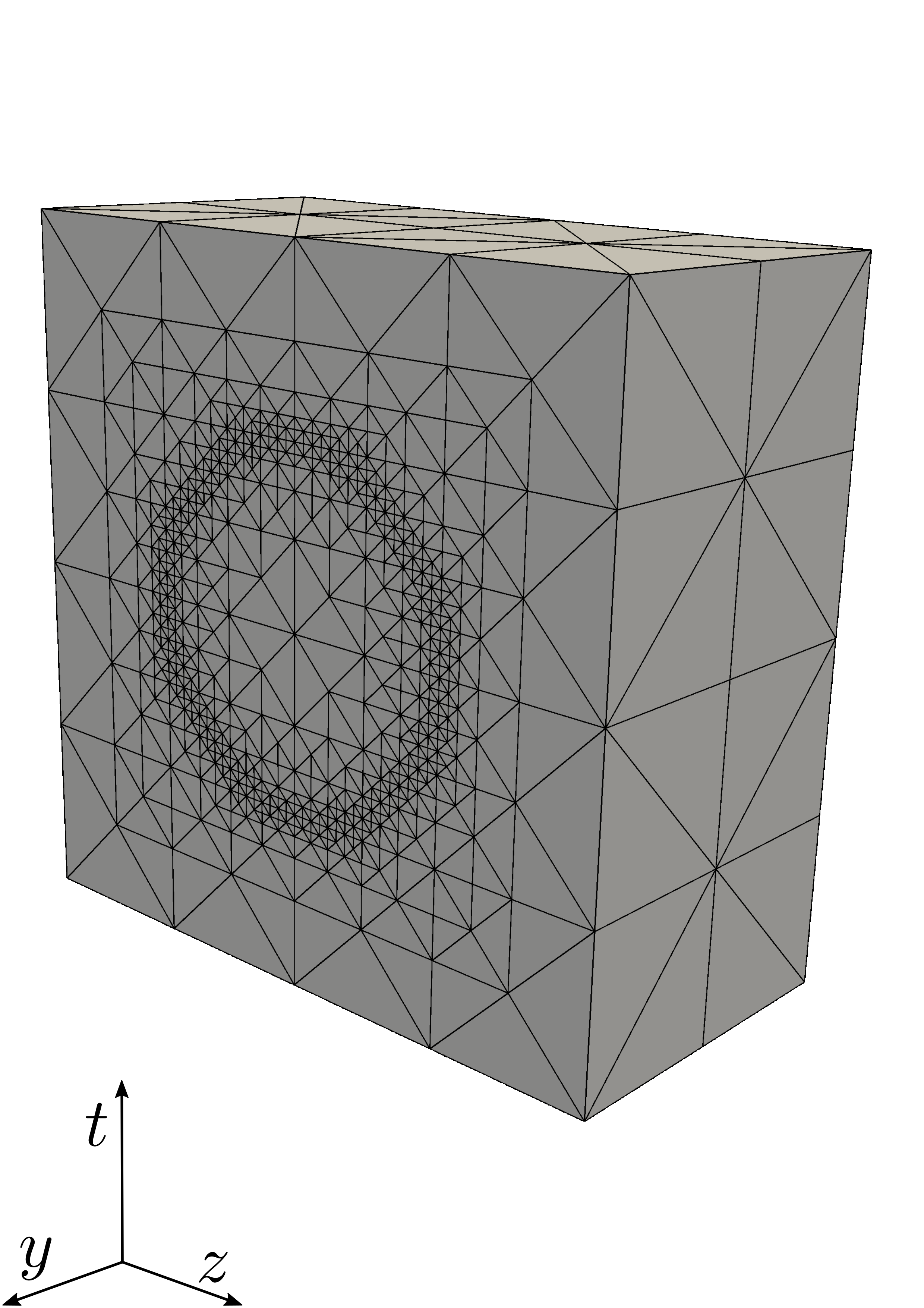}
			\label{fig:exampleMaubachArnold_4d_x}
		} 
		&
		\subfigure[]{
			\includegraphics[width=0.275\textwidth]{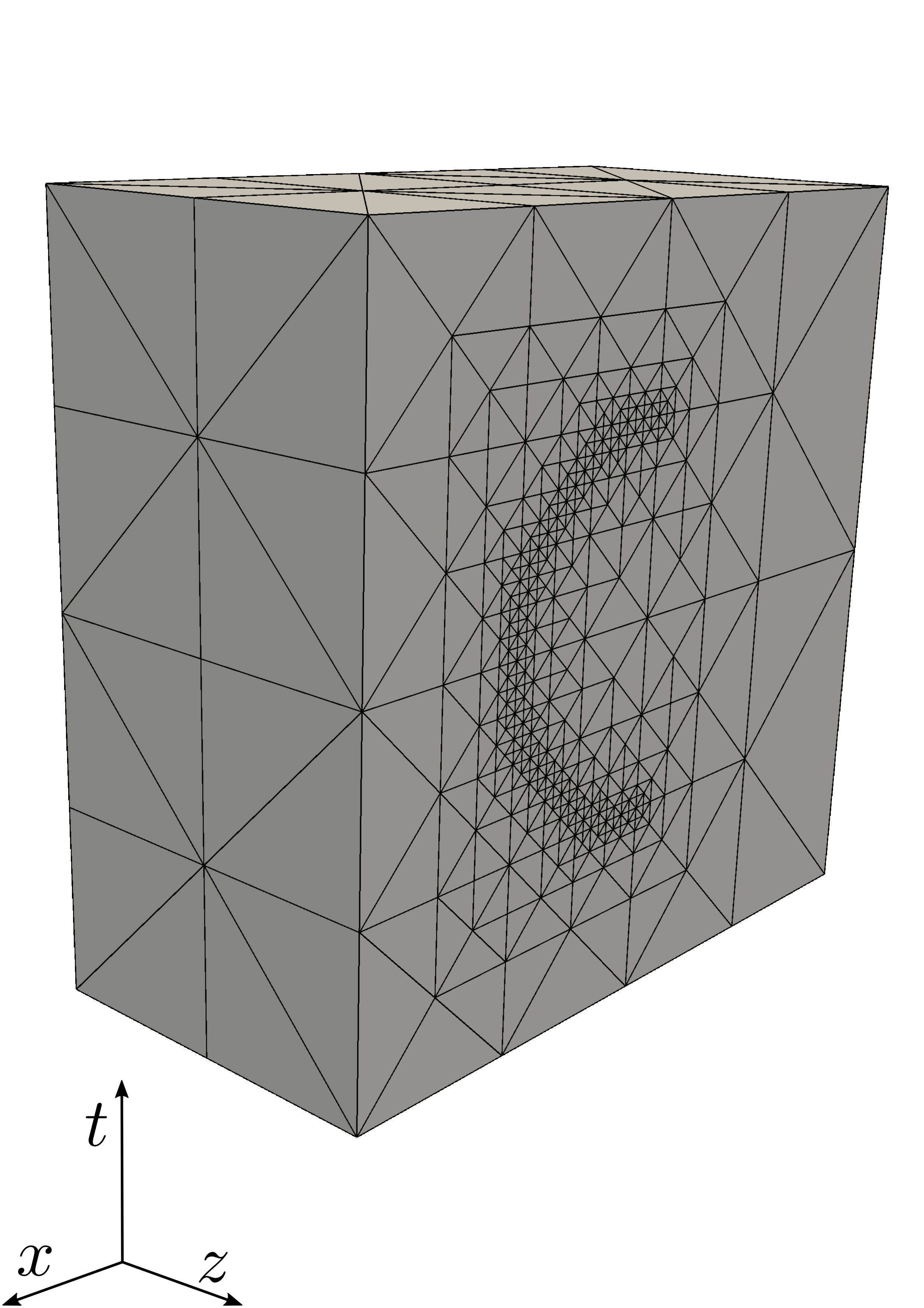}
			\label{fig:exampleMaubachArnold_4d_y}
		}
		&
		\subfigure[]{
			\includegraphics[width=0.275\textwidth]{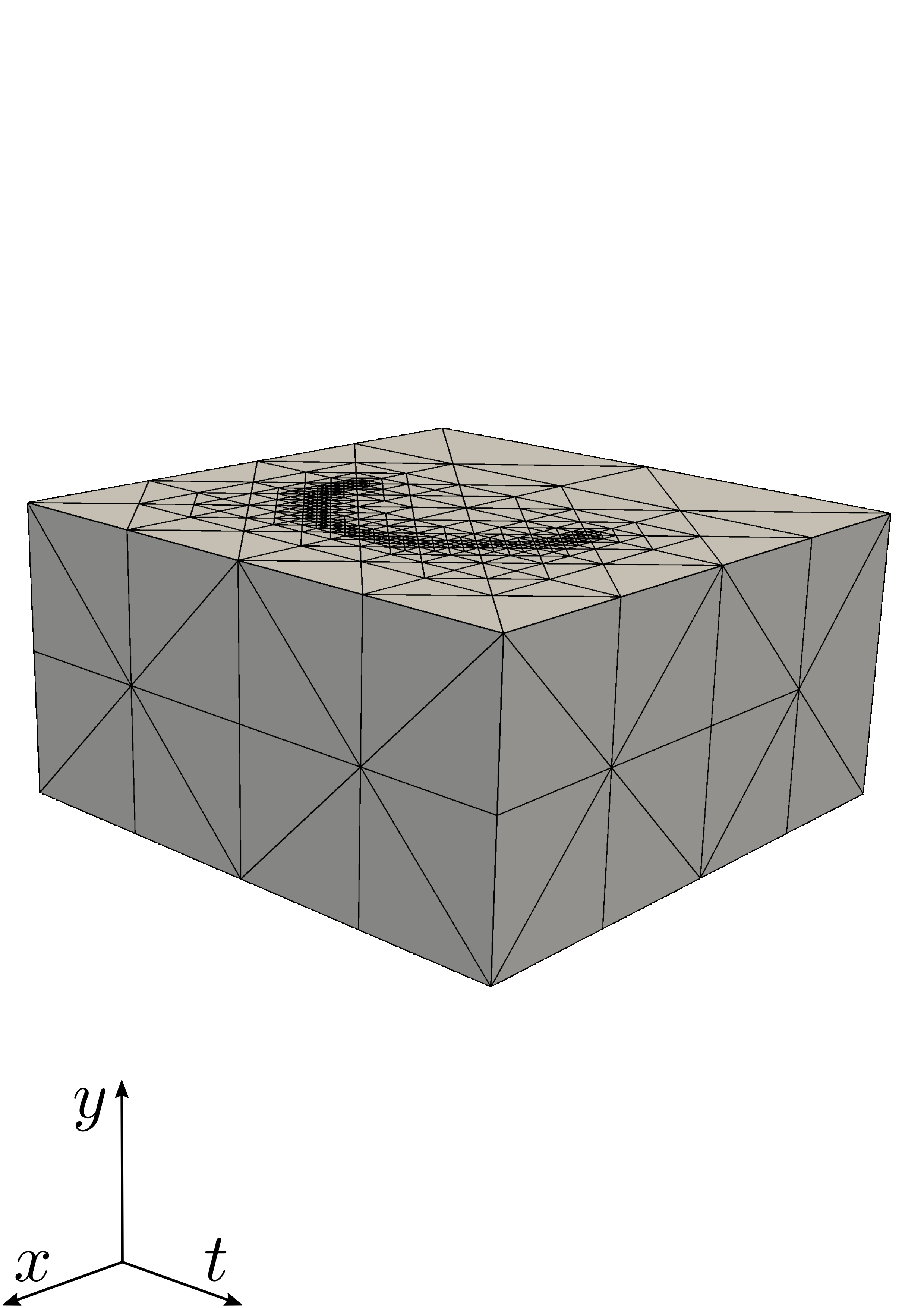}
			\label{fig:exampleMaubachArnold_4d_z}
		} 
	\end{tabular}
	\caption{Volume slices of 4-dimensional mesh \mesh{22}\ at different planes: \subref{fig:exampleMaubachArnold_4d_x} $x = 1/2$; \subref{fig:exampleMaubachArnold_4d_y} $y = 1/2$; and \subref{fig:exampleMaubachArnold_4d_z} $z = 1/2$.}
	\label{fig:exampleMaubachArnold}
\end{figure} 
Let 
\[H = \left\{ \left(x - \dfrac{1}{2}\right)^{2} + \left(y - \dfrac{1}{2}\right)^{2} + \left(z - \dfrac{1}{2}\right)^{2} + \left(t - \dfrac{1}{2}\right)^{2} = \dfrac{1}{16}, \, x \geq \dfrac{1}{2}\right\}\]
be an hemisphere of a hypersphere of radius $1/4$, centered at $(1/2,1/2,1/2,1/2)$ that is embedded in the cube $[0,1]^{4}$. We want to adapt the pentatopic mesh \mesh{0}\ to the hemisphere $H$, thus, we choose as refinement set \refinementSet{k}\ the pentatopes of \mesh{k}\ that intersect with the hemisphere $H$. That is, $\refinementSet{k} = \{\simplex{} \in \mesh{k-1} \, | \, \simplex{} \cap H \neq \emptyset\}$. 
After 22 iterations of the proposed local refinement process, the mesh \mesh{22}\ is composed of 10093008 pentatopes and 557664 vertices. In Figure \ref{fig:exampleMaubachArnold}, we make three different slices of the final mesh, \mesh{22}, with the hyperplanes $x = 1/2$, $y = 1/2$ and $z = 1/2$, see Figures \ref{fig:exampleMaubachArnold_4d_x}, \ref{fig:exampleMaubachArnold_4d_y} and \ref{fig:exampleMaubachArnold_4d_z}, respectively. We can see how the mesh has been refined locally around the hemisphere. That is, the mesh contains small elements near the hemisphere, and large elements far from the hemisphere.

\begin{figure}[t!]
	\centering
	\includegraphics[width=0.6\textwidth]{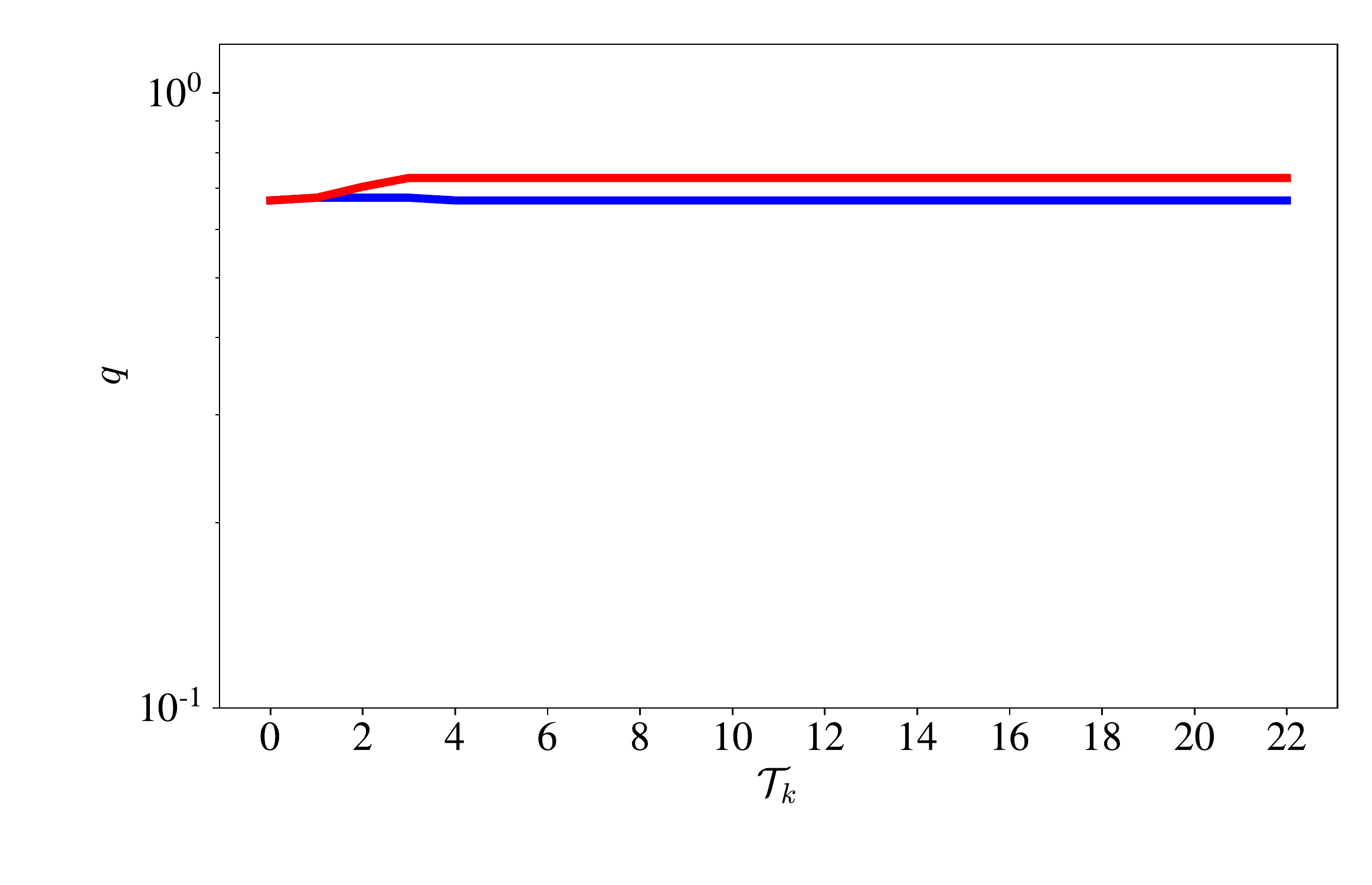}
	\caption{Evolution of the maximum (red line) and minimum (blue line) mesh quality through the mesh refinement iterations.} 
	\label{fig:qualityExampleHypercube}
\end{figure}
%
Figure \ref{fig:qualityExampleHypercube} shows the evolution of the maximum and minimum quality of the mesh during the local refinement process. We see that the maximum quality increases until iteration three, and then it remains constant. The minimum quality of the mesh decreases until iteration four, and then it stabilizes since, in posterior local refinements, the minimum mesh quality is achieved.

When we apply our co-dimensional marking process to a Coxeter-Freudenthal-Kuhn mesh, all the initial pentatopes have a bisection tree equivalent to a Maubach pentatope with tag $d = 4$.

In this example, the maximum and minimum quality are similar because the initial simplices define a Kuhn mesh. For $n$ larger than two, Kuhn meshes feature a symmetric structure that leads to only $n$ different similarity classes under successive refinement with newest vertex bisection \cite[Lemma 4.3]{maubach1995local}. Accordingly, the quality of the refined simplices lies in a tight range.

\subsection{Local refine of a 4D unstructured mesh}
\label{sec:hypersphere}

\begin{figure}[t!]
	\centering
	\begin{tabular}{cc}
		\subfigure[]{
			\includegraphics[width=0.4\textwidth]{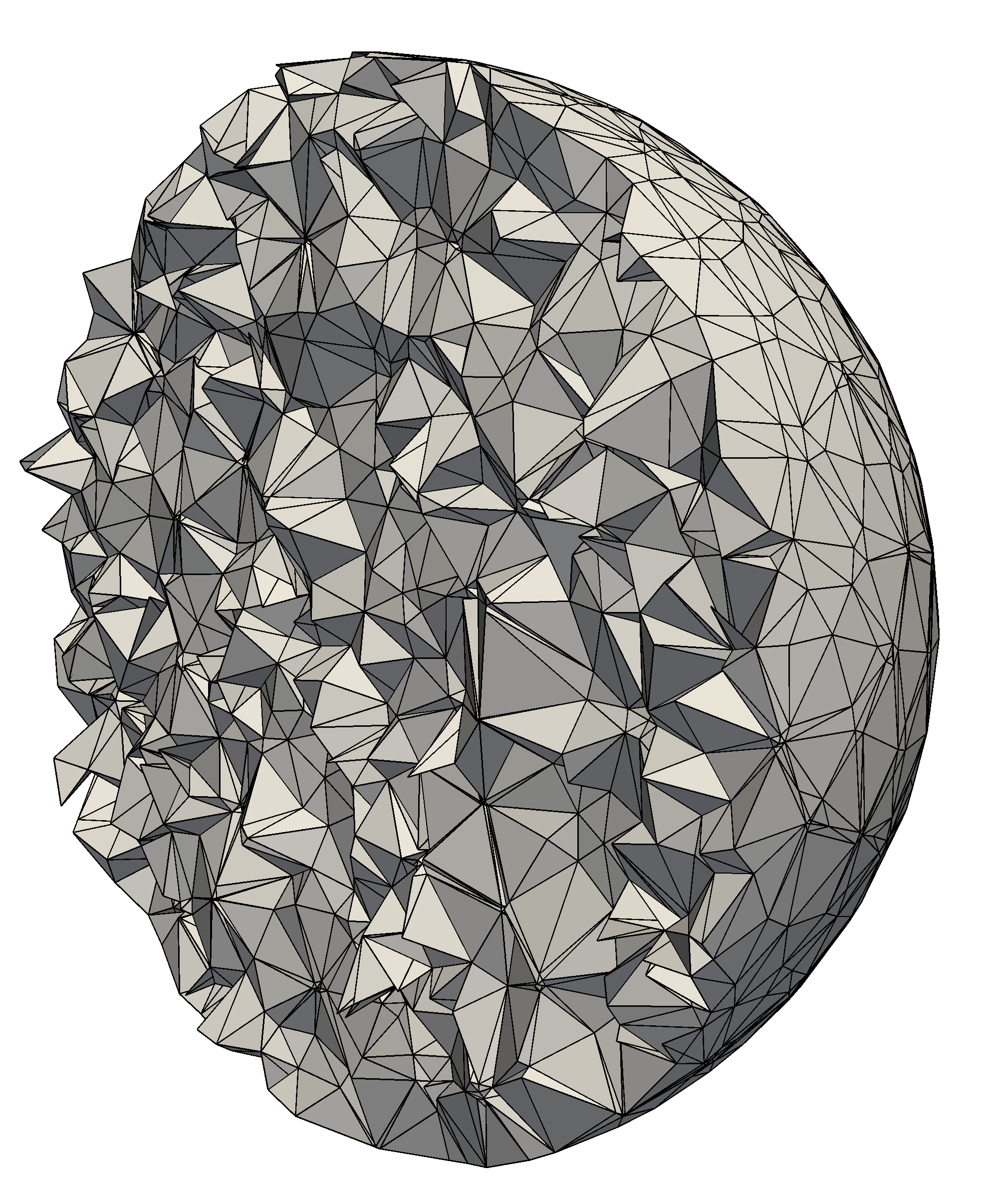}
			\label{fig:hypersphere_initial_0_t_0}
		}
		&
		\subfigure[]{
			\includegraphics[width=0.4\textwidth]{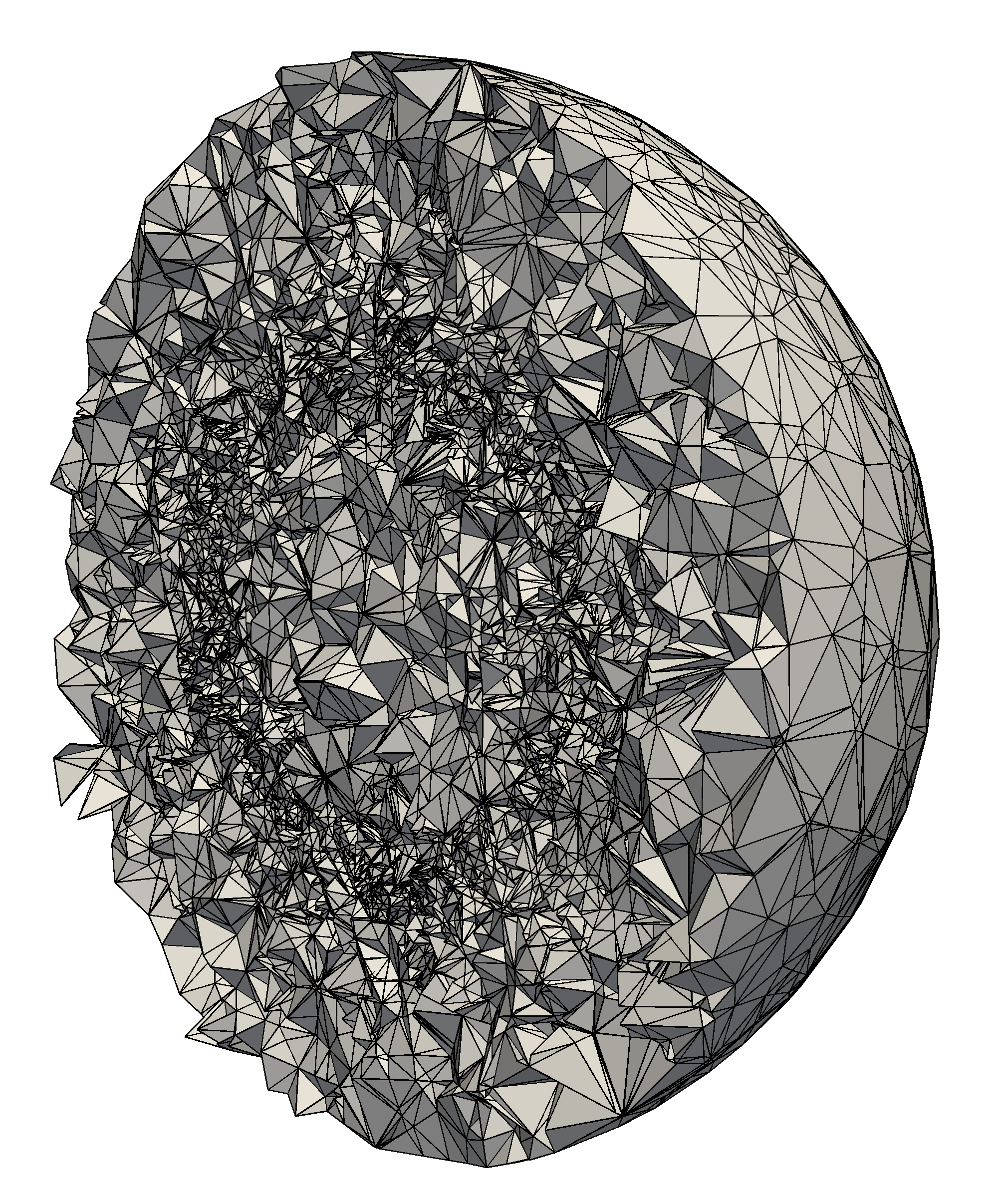}
			\label{fig:hypersphere_initial_5_t_0}
		} 
	\end{tabular}	
	\label{fig:examplehypersphere}
	\caption{Slice of the 4-simplicial mesh of an hyper-sphere with the hyper-plane $t=0$: \subref{fig:hypersphere_initial_0_t_0} initial mesh; and \subref{fig:hypersphere_initial_5_t_0} locally adapted mesh \mesh{5}.} 
\end{figure}
We show that our bisection method can be applied to locally refine unstructured simplicial meshes. In particular, to 4D unstructured pentatopic meshes. To this end, we generate an unstructured 4D mesh of an hyper-sphere of radius $1$ and centered in the origin. Then, we successively refine those elements that intersect a hyper-sphere of radius $1/2$ and centered in the origin. The initial mesh has an edge length of 0.15 and is composed of 198740 pentatopes and 10361 vertices. 
Figure \ref{fig:hypersphere_initial_0_t_0} shows a slice of \mesh{0}\ with the hyper-plane $t = 0.0$.

After 5 iterations of the refinement process, the obtained mesh \mesh{5}\ is composed of 12101892 pentatopes and 614409 vertices. We slice \mesh{5}\ with the hyper-plane $t = 0.0$ to obtain the 3D tetrahedral representation depicted in Figure \ref{fig:hypersphere_initial_5_t_0}. We can see how the mesh is refined capturing the inner hyper-sphere. The obtained results illustrate that the proposed bisection algorithm can refine unstructured simplicial meshes locally while preserving conformity. 

\begin{figure}[t!]
	\centering
	\includegraphics[width=0.6\textwidth]{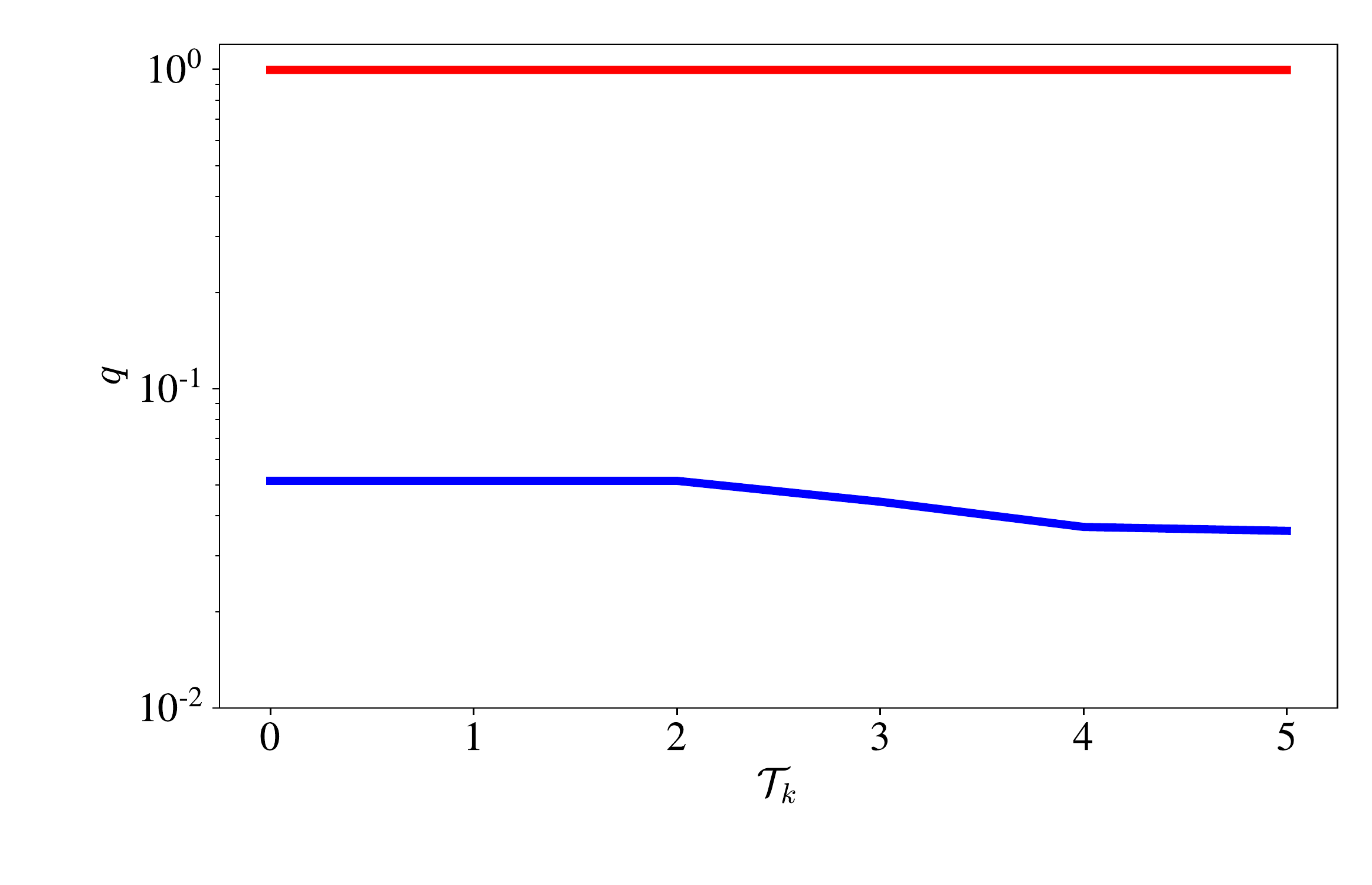}
	\caption{Evolution of the maximum (red line) and minimum (blue line) mesh quality through the mesh refinement iterations.} 
	\label{fig:qualityExamplehypersphere}
\end{figure}
Figure \ref{fig:qualityExamplehypersphere} shows the evolution of the shape quality during the refinement process. We see that the maximum quality remains constant during the refinement process. The minimum quality of the mesh decreases but does not achieve its minimum since we need to perform more local refinements.

\subsection{Local refinement in 4D space-time: time evolution of a 3D potential}

We proceed to show that we can locally refine a mesh to capture 3-di\-men\-sio\-nal manifolds that lay in 4D space-time, manifolds that result from the time evolution of an unsteady 3D potential.
We show the evolution of the gravitational potential defined by two mass particles that move along the $z$-axis. 
Let 
\begin{equation*}
	\begin{array}{c}
		V(\vec x, t)  = -G \left(\dfrac{m_{1}}{\norm{\vec{x} - \vec{p}_{1}(t)}} + \dfrac{m_{2}}{\norm{\vec{x} - \vec{p}_{2}(t)}}\right) \\
		\\
		\vec{p}_{1}(t) =  \vec{p}_{1} + (0,0,vt),\, t \in [0,1] \\
		\vec{p}_{2}(t) =  \vec{p}_{2} - (0,0,vt),\, t \in [0,1]
	\end{array}
\end{equation*}
the equation that defines the gravitational potential. 
For a given iso-value $V_{0}$, $V(\vec{x}, t) = V_{0}$ defines a 3D embedded manifold in 4D space. Let $H$ be the hyper-cylinder with spherical basis defined by the equations
\begin{equation*}
	\begin{array}{c}
		\left(x-\dfrac{1}{2}\right)^2+\left(y-\dfrac{1}{2}\right)^2+\left(z-\dfrac{1}{2}\right)^2 = 1
		\\
		-0.1 \leq t \leq 1.1.
	\end{array}
\end{equation*}
In this example, we choose the iso-value $V_{0} = -10$ and the parameters $G = 1$, $m_{1} = 1$, $m_{2} = 1$, $\vec{p}_{1} = (1/2, 1/2, 1/8)$, $\vec{p}_{2} = (1/2, 1/2, 7/8)$ and $v = 3/8$. 

We generate an adapted pentatopic mesh by locally refining an initial mesh around the manifold. We generate the initial mesh \mesh{0}\ composed of 7345 pentatopes and 576 vertices. We generate the set of pentatopes that intersect $H$, $F_{k} = \{ \simplex{} \in \mesh{k-1} \,|\, \simplex{} \cap H \neq \emptyset\}$. Then, for each pentatope in $F_{k}$ we compute the curvature of $V(\vec{x}, t)$ at each simplex using the formula $$\edge{\simplex{}} = \sum_{i=0}^{4}\left| h_{i}^{T}\nabla^{2}V(\vec{x}_{i}, t_{i})h_{i} \right|,$$ where $\nabla^{2}V(\vec{x}_{i}, t_{i})$ is the Hessian matrix of the potential $V(\vec{x},t)$ evaluated at the vertices $(\vec{x}_{i}, t_{i})$ of \simplex{},  and $h_{i} = (\vec{x}_{i},t_{i}) - c_{M}$, where $c_{M}$ is the center of mass of \simplex{}. After that, we choose as refinement set $\refinementSet{k}$ the 10\% of the pentatopes of $F_{k}$ with more curvature. The idea is to adapt the pentatopic mesh not only to the elements that intersect the iso-surface but also to the areas of the iso-surface with more curvature. 

\begin{figure}[t!]
	\centering
	\begin{tabular}{cc}
		\subfigure[]{
			\includegraphics[width=0.33\textwidth]{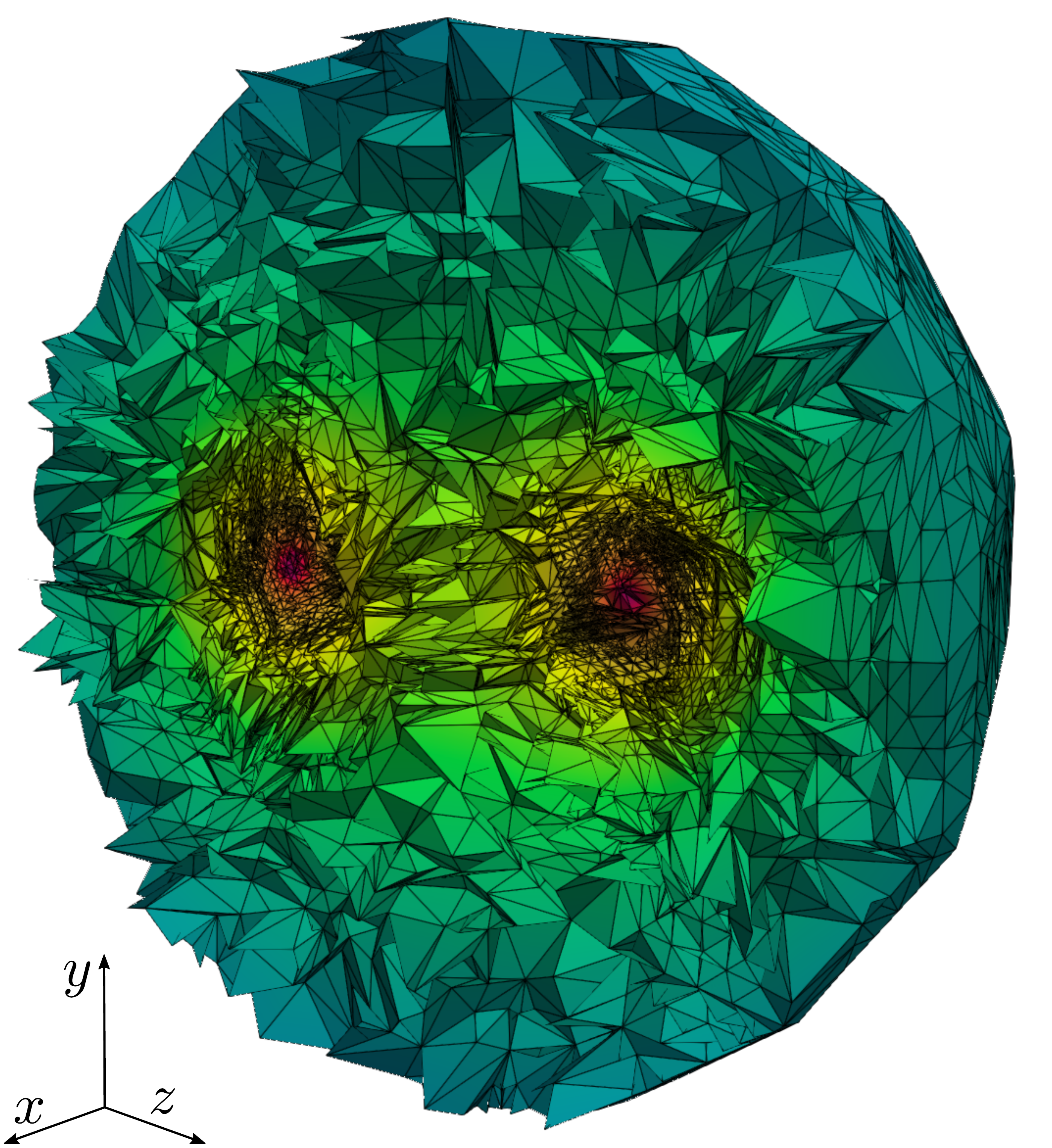}
			\label{fig:3dmesh1}
		} 
		&
		\subfigure[]{
			\includegraphics[width=0.33\textwidth]{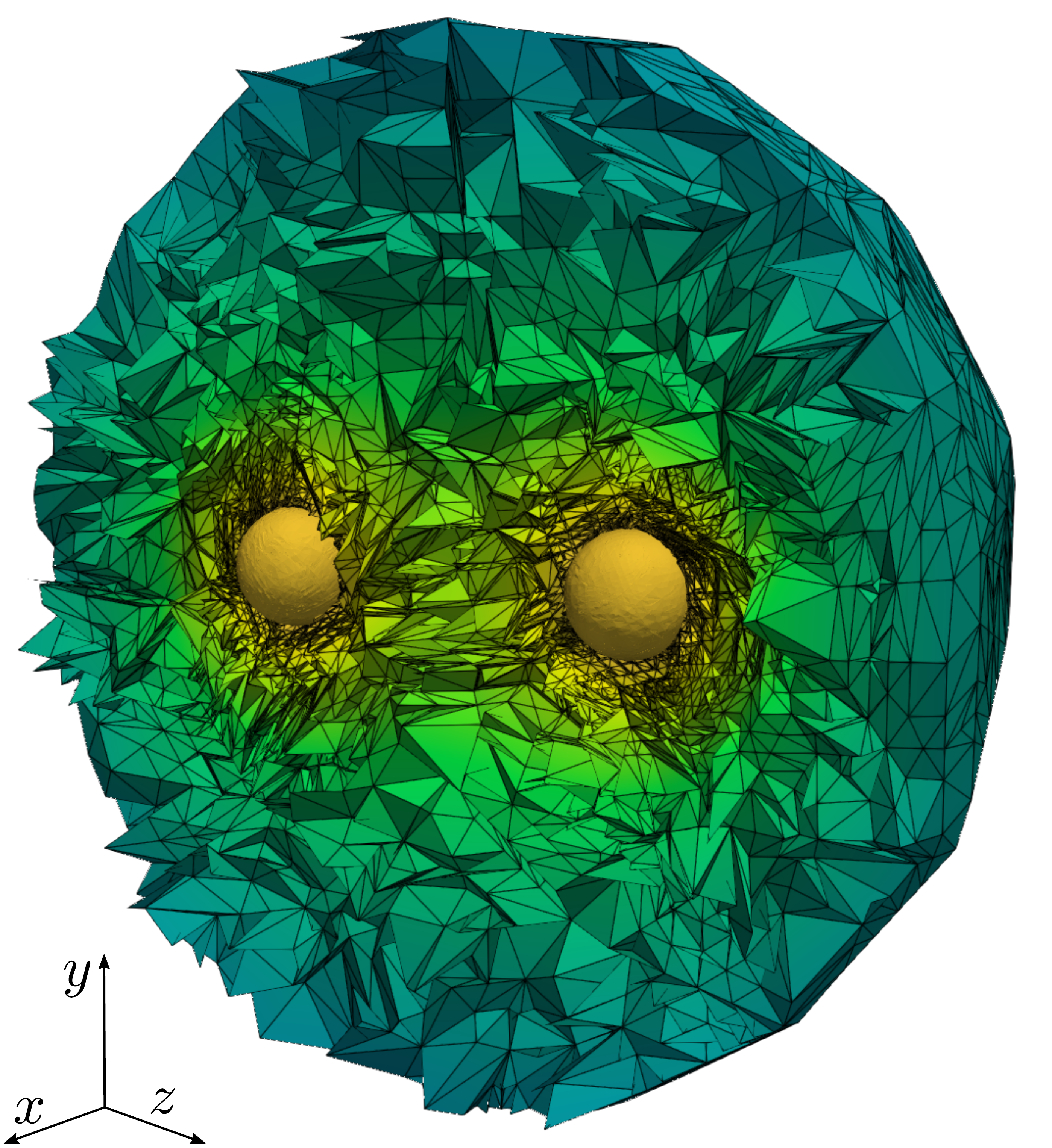}
			\label{fig:contour1}
		} 
		\\
		\subfigure[]{
			\includegraphics[width=0.33\textwidth]{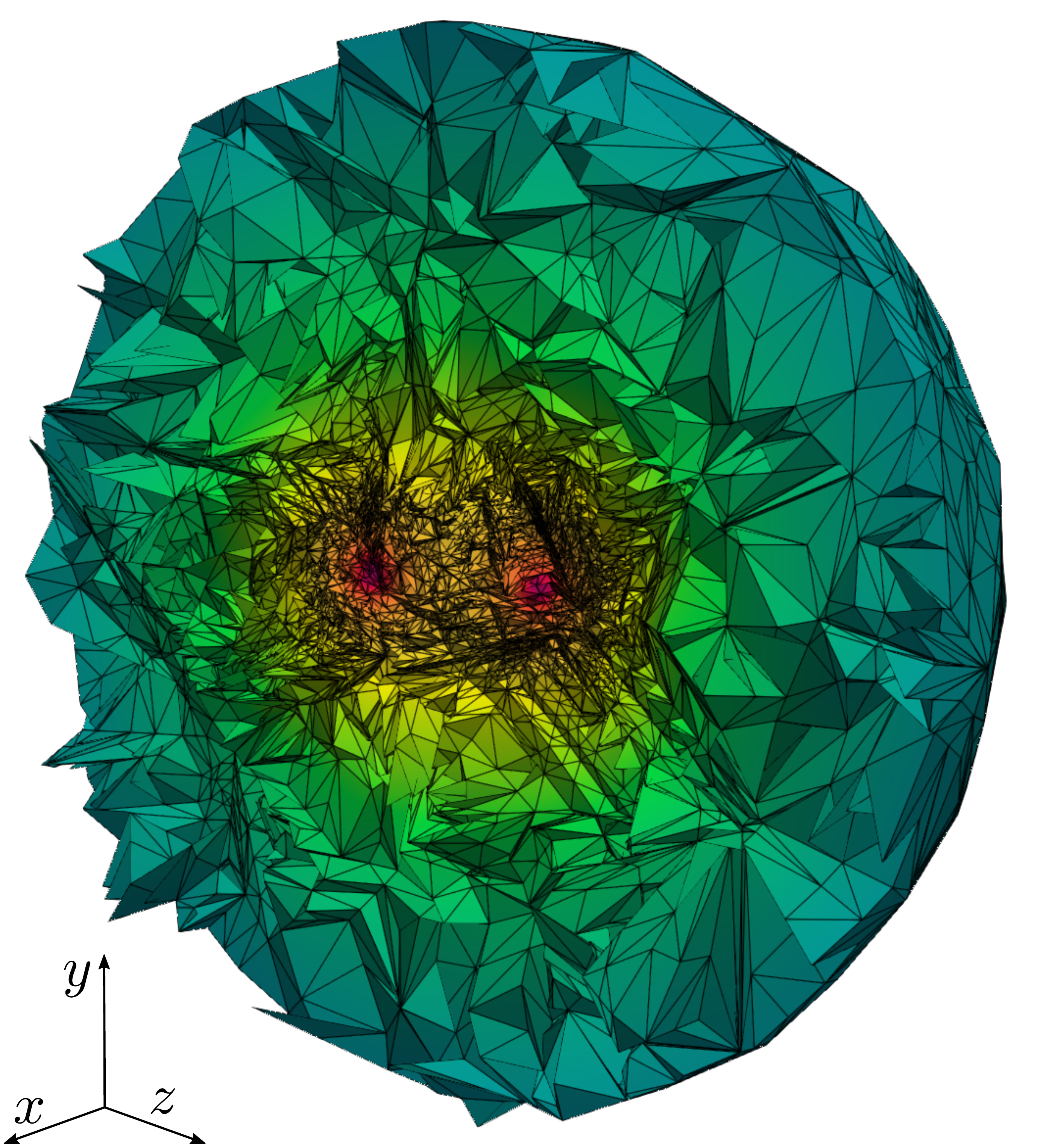}
			\label{fig:3dmesh2}
		} 
		&
		\subfigure[]{
			\includegraphics[width=0.33\textwidth]{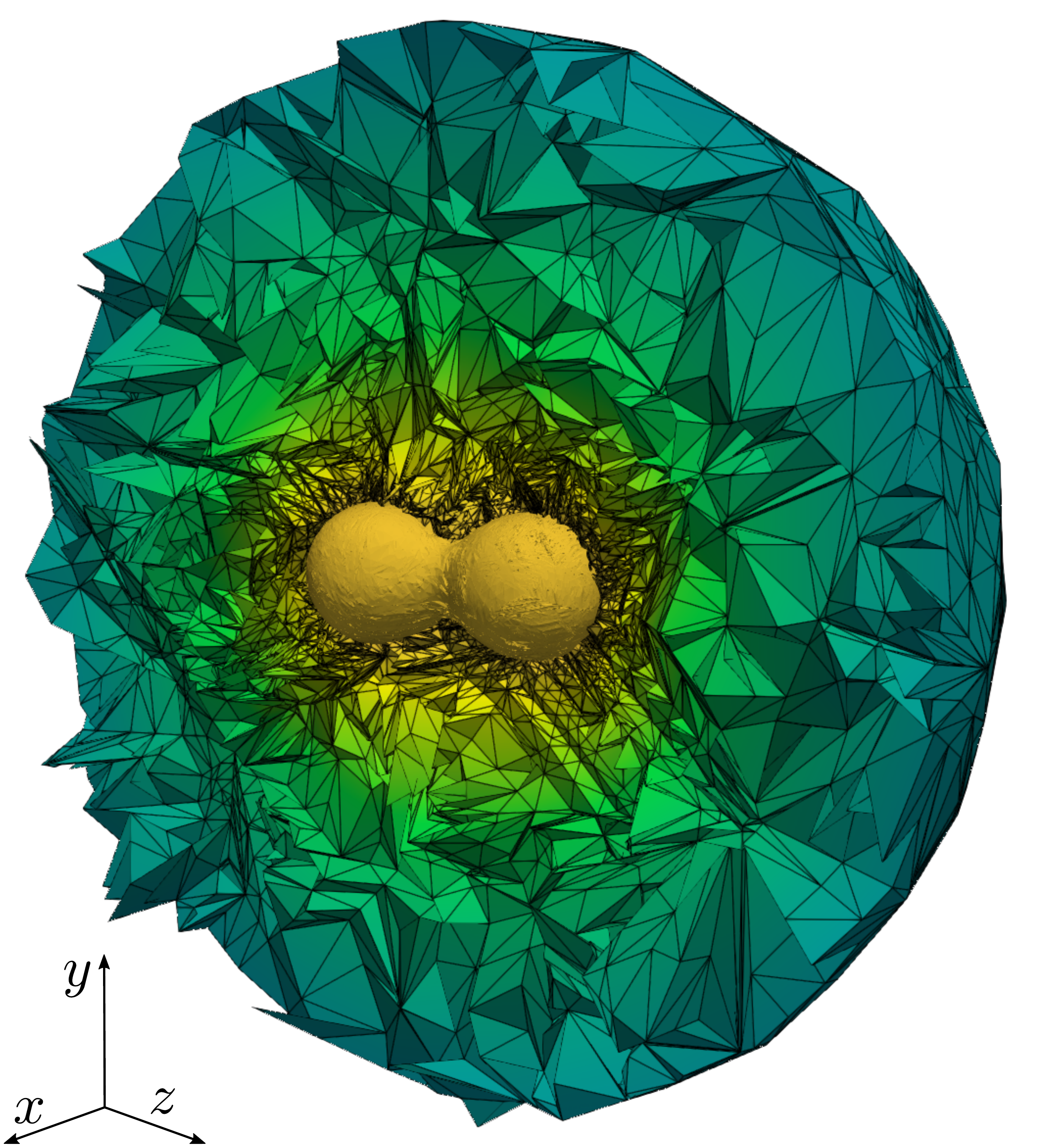}
			\label{fig:contour2}
		} 
		\\
		\subfigure[]{
			\includegraphics[width=0.33\textwidth]{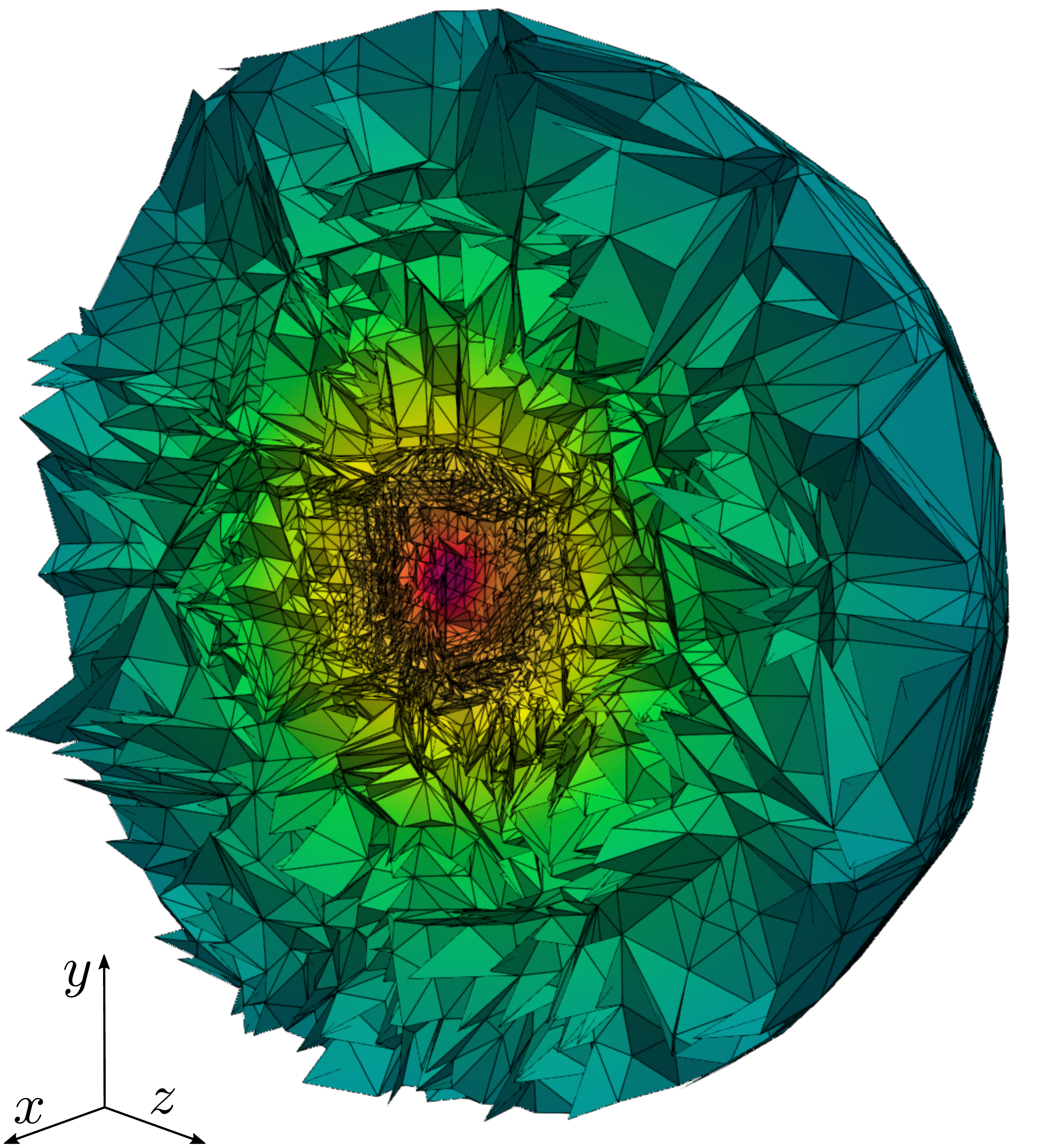}
			\label{fig:3dmesh3}
		}
		
		&
		\subfigure[]{
			\includegraphics[width=0.33\textwidth]{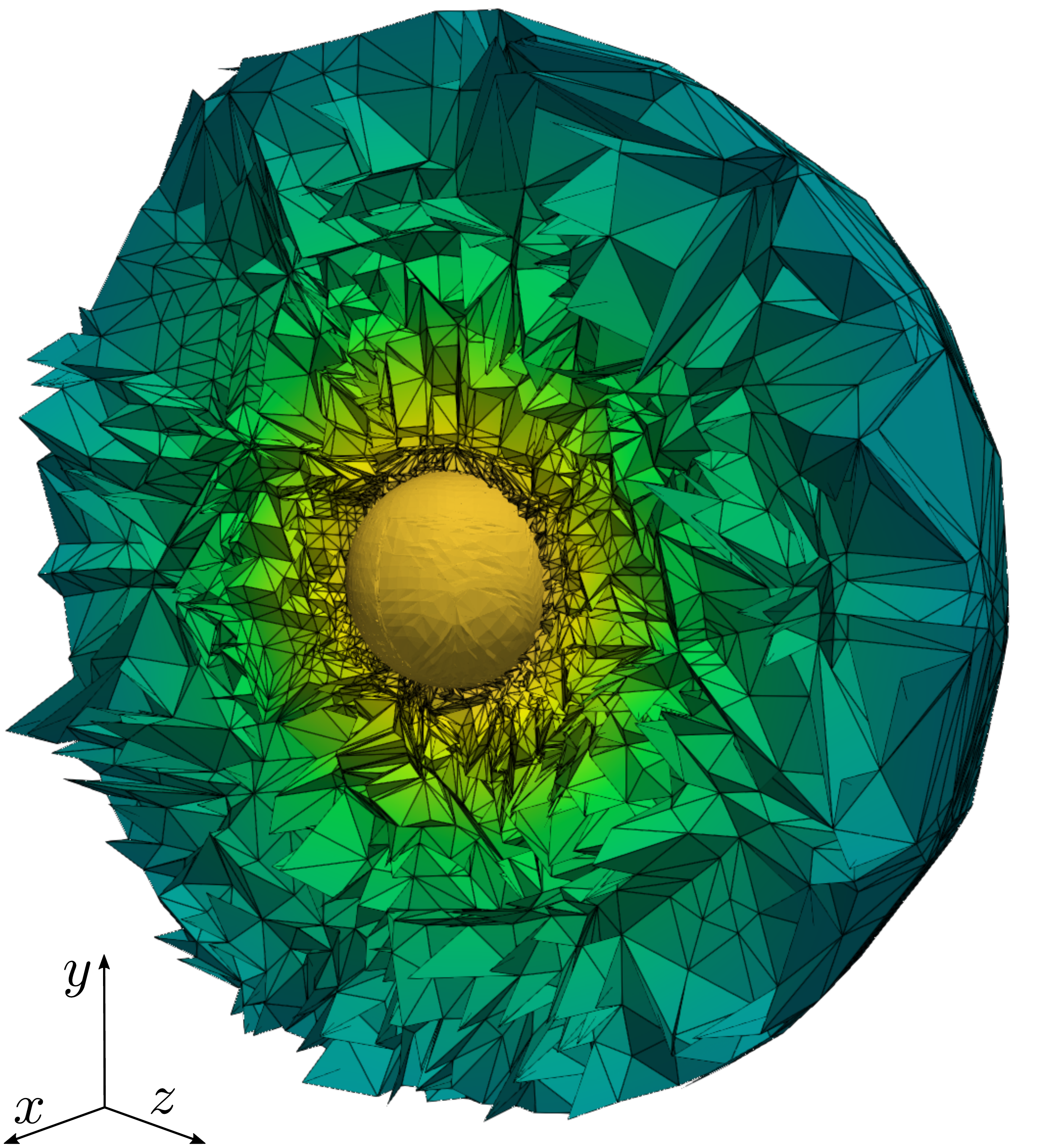}
			\label{fig:contour3}
		}
	\end{tabular}
	\caption{Volume slices of \mesh{18}\ at different time instants colored by potential value. In columns, slice of the mesh (a,c,e) without and (b,d,f) with the iso-potential manifold. In rows, slices with: (a,b) $t = 0.0$; (c,d) $t = 0.5$; and (e,f) $t=1.0$.}
	\label{fig:examplePotentialT}
\end{figure} 
\begin{figure}[t!]
	\centering
	\begin{tabular}{cc}
		\subfigure[]{
			\includegraphics[width=0.35\textwidth]{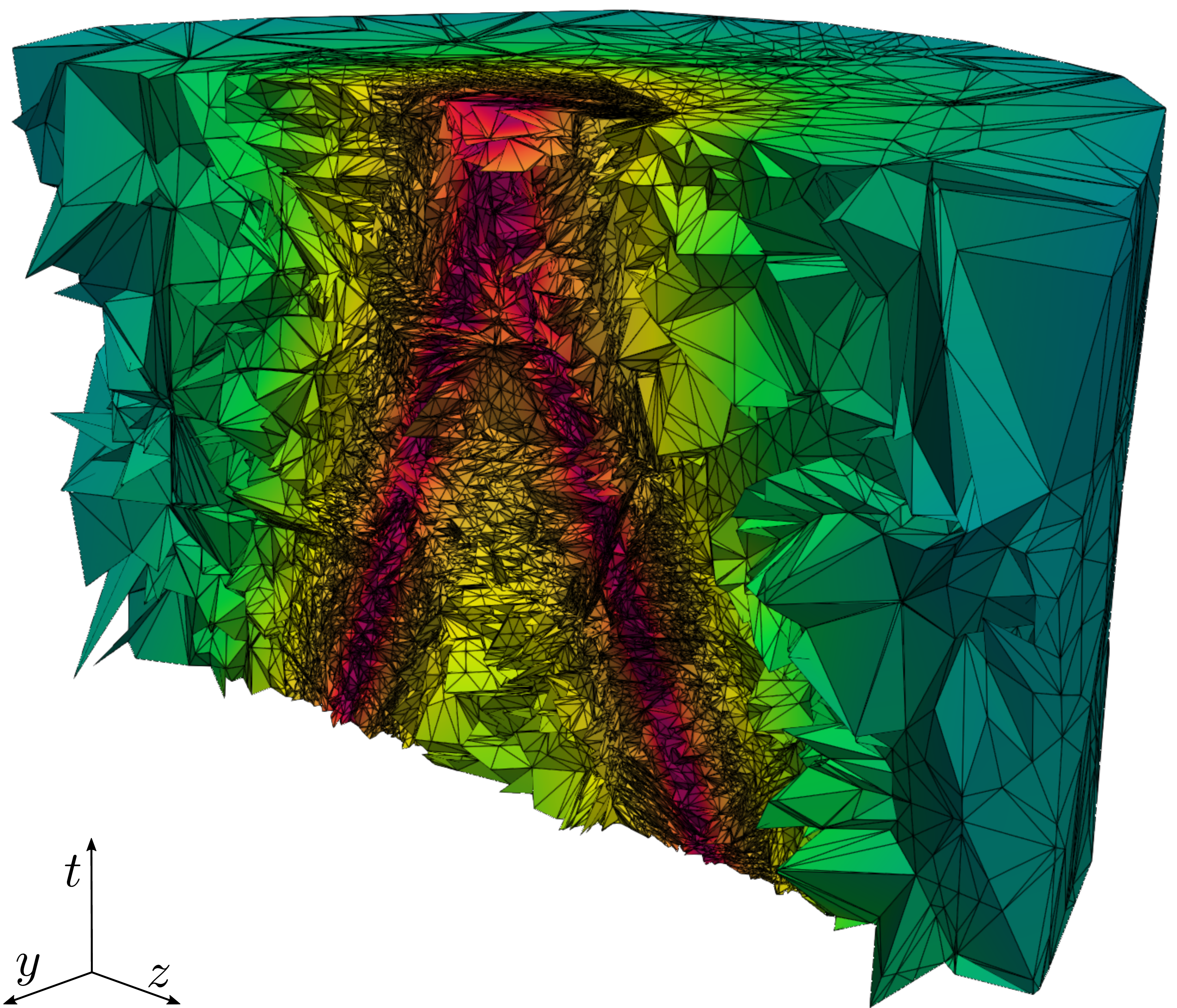}	
			\label{fig:3dmesh4}
		} 
		&
		\subfigure[]{
			\includegraphics[width=0.35\textwidth]{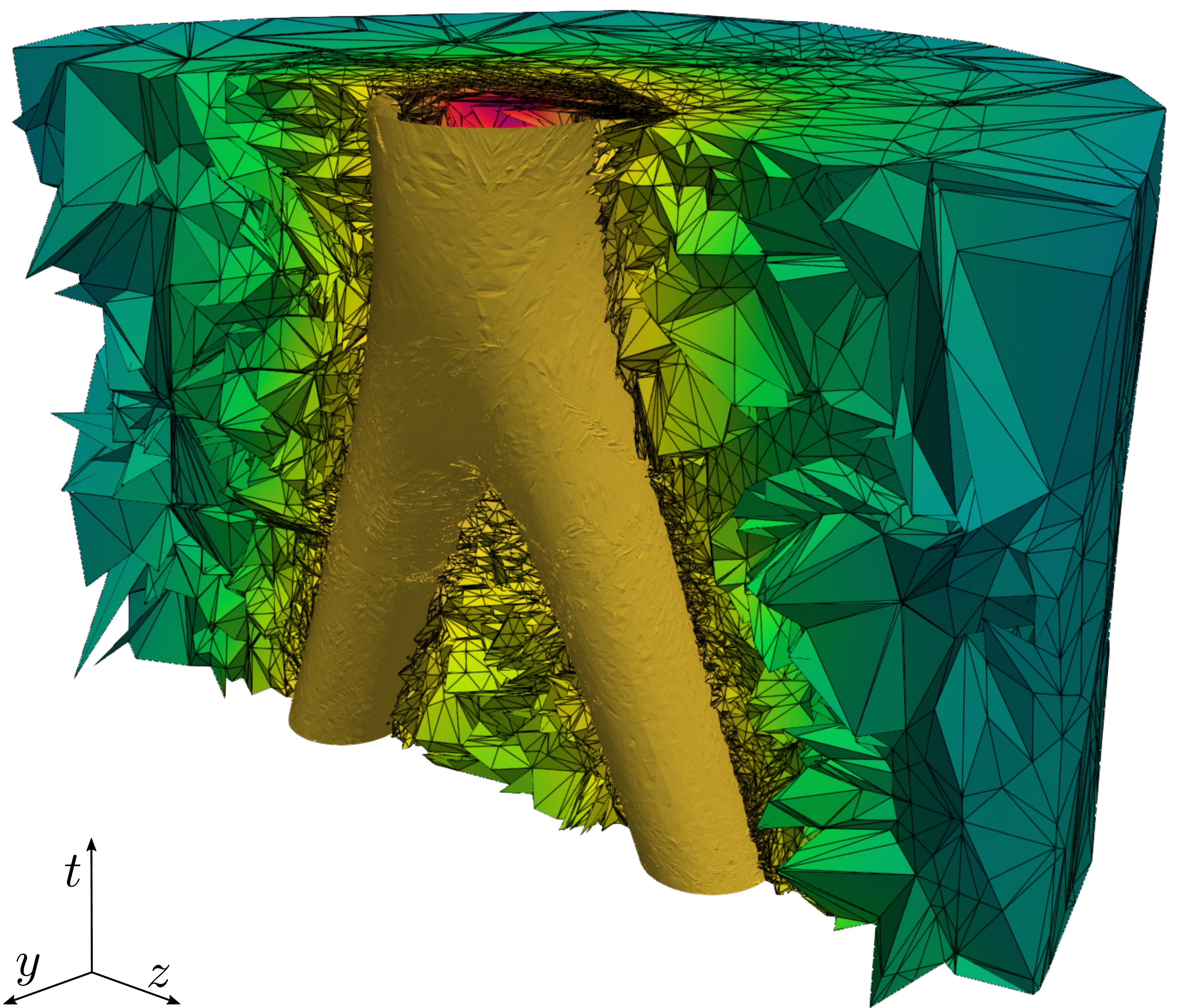}
			\label{fig:contour4}
		}
	\end{tabular}
	\caption{Volume slice with the hyper-plane $x = 1/2$. In Figures \subref{fig:3dmesh4} and \subref{fig:contour4} we obtain the 3D space-time mesh $(z,y,t)$, where we can see the time evolution of the iso-surface defined by the gravitational potential. We can see how the mesh is adapted to capture the movement of the two particles.
	}
	\label{fig:examplePotentialX}
\end{figure} 
After 18 iterations of the local refinement process, the generated mesh \mesh{18}\ has 12115582 pentatopes and 619571 vertices. To visualize the obtained mesh, we sliced it with a hyper-plane to obtain a 3D tetrahedral representation. Figures \ref{fig:3dmesh1}, \ref{fig:3dmesh2}, and \ref{fig:3dmesh3} show a slice of the pentatopic mesh with the hyper-planes $t = 0$, $t = 0.5$ and $t = 1$, respectively. The mesh has been locally refined around the iso-surface and therefore, we have smaller elements near the iso-surface and large elements far from the iso-surface. Figures \ref{fig:contour1}, \ref{fig:contour2}, and \ref{fig:contour3} show the iso-surface that is extracted from the mesh. These slices show that the iso-surface starts as two different connected components and then it merges as one connected component. Figure \ref{fig:3dmesh4} shows a slice of the pentatopic mesh with the hyper-plane $x = 0.5$, generating the space-time mesh $(z,y,t)$. We can see how the mesh captures the time evolution of the iso-surface defined by $V(\vec{x},t)$. Figure \ref{fig:contour4} shows the iso-surface that is extracted from the space-time mesh. 

\begin{figure}[t!]
	\centering
	\includegraphics[width=0.6\textwidth]{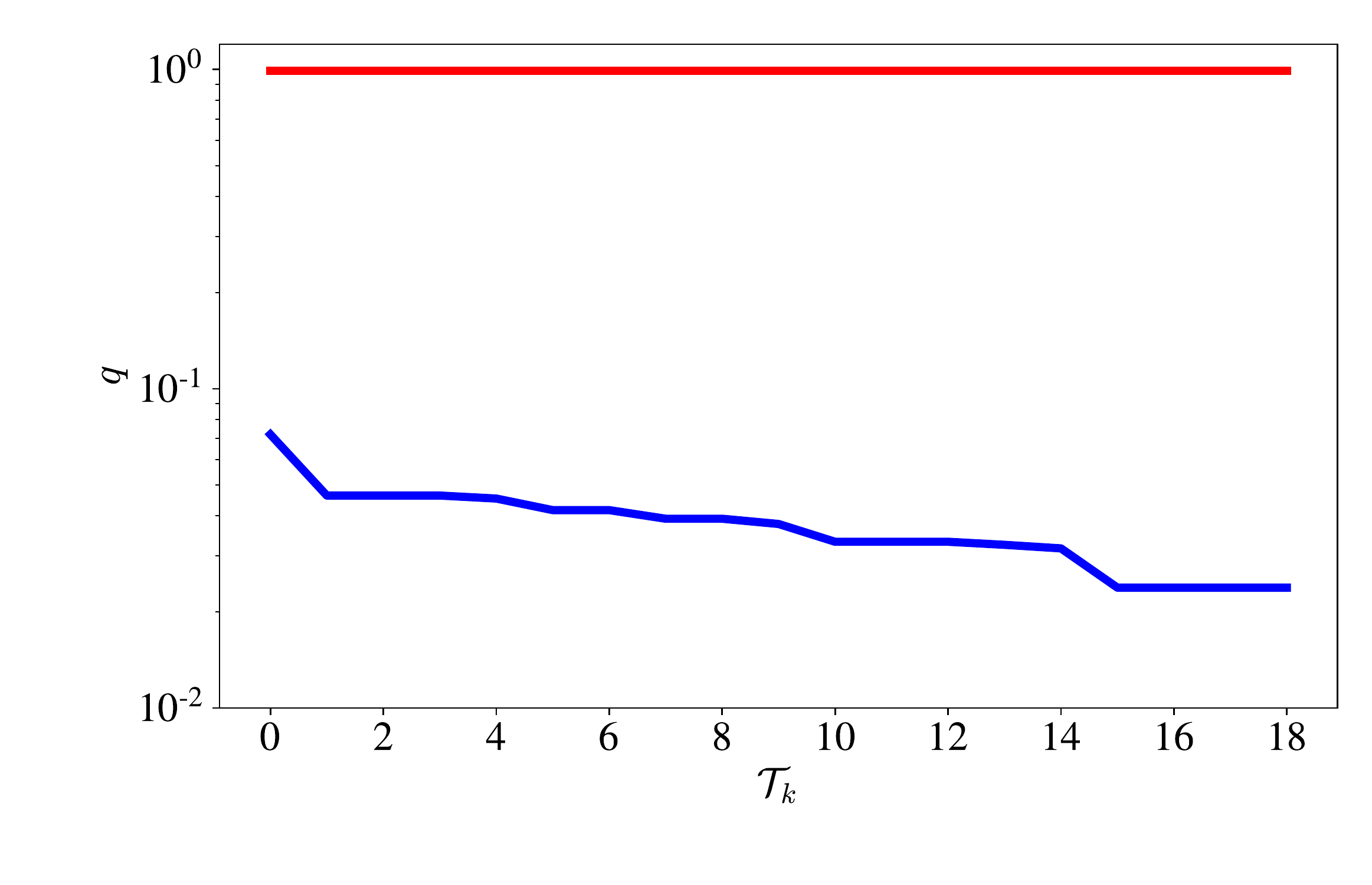}
	\caption{Evolution of the maximum (red line) and minimum (blue line) mesh quality through the mesh refinement iteration.} 
	\label{fig:qualityExampleGravitational}
\end{figure}
Figure \ref{fig:qualityExampleGravitational} shows the evolution of the maximum and minimum quality of the mesh during the local refinement process. We see that the maximum quality remains constant during the refinement process. The minimum quality of the mesh decreases until iteration 16, and then it stabilizes since, in posterior local refinements, the minimum mesh quality is achieved.

\section{Concluding remarks}
\label{sec:conclusions}

To heuristically enforce a bisection method that almost fulfills the similarity and locality properties of newest vertex bisection, we have considered two key ingredients. First, independently for each simplex, the marked bisection is active only in the first $n$ bisection steps. Therefore, it is responsible for a slight initial increase in the number of similarity classes and the cost of completing the conformal closure. Second, the marking process and the two stages heuristically enforce the condition that after successive $n$ uniform marked bisections, we obtain a conformal and reflected mesh. According to \cite{stevenson2008completion}, these are sufficient conditions to switch to newest vertex bisection.

In conclusion, we have proposed and implemented the first practical $n$-dimensional multi-stage bisection for unstructured conformal meshes. In future work, we will prove that the proposed method features the properties of a practical bisection method.

In perspective, guaranteeing unstructured conformal refinement would enable adaptive applications on $n$-dimensional complex geometry. In this case, the complexity of the geometry would be handled by the flexibility of unstructured conformal meshes. 
Furthermore, on these meshes, the refinement would almost fulfill the similarity and locality properties of pure newest vertex bisection for $n$-simplicial adaption.

\section{Acknowledgements}
\label{sec:acknowledgements}

This project has received funding from the European Research Council (ERC) under the European Union's Horizon 2020 research and innovation programme under grant agreement No 715546. This work has also received funding from the Generalitat de Catalunya under grant number 2017 SGR 1731. The work of Xevi Roca has been partially supported by the Spanish Ministerio de Econom\'ia y Competitividad under the personal grant agreement RYC-2015-01633.

\bibliographystyle{elsarticle-num-names}
\bibliography{references.bib}

\appendix

\section{Checking of conformity and reflectivity}
\label{sec:checkings}

\subsection{Conformity check}
\label{sec:conformity}
The main goal of this appendix is to propose an algorithm to check the conformity of the meshes generated by the local refinement algorithm. The main idea of the algorithm is to use the multi-indices of the nodes of the original and bisected meshes.

Stevenson states in \cite{stevenson2008completion} a definition for conformal meshes slightly different from the standard one. Specifically, he considers that a simplicial mesh is \emph{conformal} if the following two conditions are satisfied:
\begin{enumerate}
	\item[(C1)] For any $\simplex{} \in \mesh{}$, $\simplex{} \cap \partial \Omega$ is the union of entities of \simplex{}.
	\item[(C2)] Each $x \in \simplex{1} \cap \simplex{2}$, with $\simplex{1}, \simplex{2} \in \mesh{}$, for which any open ball $B \ni x $, any $y \in \simplex{1} \cap B \cap\Omega$, $y^{\prime} \in  \simplex{2} \cap B \cap\Omega$ are connected by a path through $B \cap \Omega$, lies on a joint $k$-entity of \simplex{1}\ and \simplex{2}. 
\end{enumerate} 
When $\Omega$ nowhere lies simultaneously on both sides of an $(n-1)$-
dimensional part of its boundary, (C2) is equivalent to the standard definition of conformity, see Section \ref{sec:preliminaries}. If, in addition, $\partial \Omega$ is everywhere $(n-1)$-dimensional, i.e., if $\Omega = \textnormal{int} (\overline{\Omega})$, then (C2) implies (C1).

With that definition Stevenson \cite[Theorem 3.1]{stevenson2008completion} states that if a mesh \mesh{}\ fulfills the condition (C2) then, it is enough to check the conformity of \mesh{}\ only through faces. We rewrite \cite[Theorem 3.1]{stevenson2008completion} as follows:
\begin{theorem}
	\label{thm:stevenson}
	A mesh \mesh{}\ of $\Omega$ satisfies (C2) if and only if any $\simplex{1}, \simplex{2} \in \mesh{}$, for which $\simplex{1} \cap \simplex{2}\cap\Omega$ contains a point interior to a face of \simplex{1}, or \simplex{2}, are neighbors.
\end{theorem}
Specifically, if a mesh \mesh{}\ is conformal by faces, then Theorem \ref{thm:stevenson} ensures automatically that \mesh{}\ is conformal by all the $k$-entities, where $1 \leq k \leq n-2$. Thus, Theorem \ref{thm:stevenson} ensures that checking the conformity of the faces automatically checks the conformity of all the entities of the mesh.

Let \mesh{0}\ be a conformal simplicial mesh and \refinementSet{0}\ the set of simplices to bisect. Consider the mesh \mesh{1}\ obtained after refining the set of simplices \refinementSet{0}\ of \mesh{0}\ using the local refinement algorithm. Since we bisect edges during the refinement process, we have two types of multi-indices in \mesh{1}: the original vertices \vertex{i}\ of \mesh{0}, and the new vertices $[v_{i_{1}}, v_{i_{2}},\ldots ,v_{i_{k}}]$, which are generated during the bisection process. We know that \mesh{0}\ is conformal, and we want to check that \mesh{1}\ is conformal, too.

The idea is devised in two steps. The first one is to check that there are no faces shared by more than two simplices. The second one is to check that all the boundary faces of \mesh{1}\ are contained in the boundary faces of \mesh{0}. The boundary check allows us to ensure that we do not create holes during the bisection process. For that purpose, we use a dictionary data structure to define a map between a sorted face \face{}\ and the simplices that contain it. We propose to create two dictionaries: a dictionary of inner faces, \innerDict{}; and a dictionary of boundary faces \boundaryDict{}. Thus, for a given sorted face \face{}, the operation $\innerDict{}[\face{}]$ and $\boundaryDict{}[\face{}]$ return the simplices, or simplex, that contain the sorted face \face{}.

\begin{algorithm}[t!]
	\caption{Generation of a dictionary of faces}
	\label{alg:getDictionaryOfFaces}
	\begin{algorithmic}[1]
		\Require{\type{Mesh} \mesh{}}
		\Ensure{\type{Dictionary} \innerDict{}, \type{Dictionary} \boundaryDict{}}
		\Function{getFaces}{\mesh{}}
		\State $\facesDict{} = \{\};\, \innerDict{} = \{\};\, \boundaryDict{} = \{\}$ \Comment{Dictionaries of all the faces, inner faces and boundary faces, respectively.}
		\For{$\simplex{}$ \textbf{in} $\mesh{}$}
		\label{line:loopSimplicesDictionary}
		\State $(\bvertex{i_{0}}, \bvertex{i_{1}}, \ldots, \bvertex{i_{n}}) = \Call{sort}{\simplex{}}$
		\label{line:sortSimplicesLexi}
		\For{$j = 0, 1,\ldots, n$}
		\State $\face{j} = \Call{oppositeFace}{(\bvertex{i_{0}}, \bvertex{i_{1}}, \ldots, \bvertex{i_{n}}), \bvertex{i_{j}}}$ 
		\label{line:oppositeFacesDictionary}
		\If{$\facesDict{}[\face{j}] = \emptyset$}
		\State $\facesDict{}[\face{j}] = \simplex{}$
		\label{line:assignOneDictionary}
		\Else
		\State $\facesDict{}[\face{j}] = \Call{append}{\facesDict{}[\face{j}], \simplex{}}$
		\label{line:assignTwoDictionary}
		\EndIf
		\EndFor
		\EndFor
		\For{$\face{}$ \textbf{in} $\texttt{keys}(\facesDict{})$}
		\label{line:typeOneFaceDictionary}
		\If{$\Call{length}{\facesDict{}[\face{}]} = 1$}
		\State $\boundaryDict{}[\face{}] = \facesDict{}[\face{}]$
		\ElsIf {$\Call{length}{\facesDict{}[\face{}]} = 2$}
		\State $\innerDict{}[\face{}] = \facesDict{}[\face{}]$
		\Else
		\State \textbf{Error}: A face cannot be shared by more than two simplices.
		\EndIf
		\EndFor
		\label{line:typeTwoFaceDictionary}
		\State \Return \innerDict{}, \boundaryDict{}
		\EndFunction
	\end{algorithmic}
\end{algorithm}
The proposed method to generate \innerDict{}\ and \boundaryDict{}\ is shown in Algorithm \ref{alg:getDictionaryOfFaces}. First, we do a loop over the simplices of the mesh \mesh{}, see Line \ref{line:loopSimplicesDictionary}, and sort them using the lexicographic order,  see Line \ref{line:sortSimplicesLexi}. Then, we compute the faces of the sorted simplex using the opposite faces method, a procedure that for a given simplex \simplex{}\ and a given vertex \vertex{}, returns the opposite face \face{}\ to the vertex \vertex{}\ inside the simplex \simplex{}, see Line \ref{line:oppositeFacesDictionary}. Since the vertices of the simplex are ordered, the vertices of each face are also ordered. After that, if the sorted face \face{}\ is not inside the dictionary \facesDict{}, we add it and assign as a value the simplex \simplex{}, see Line \ref{line:assignOneDictionary}. If the sorted face \face{}\ is already in the dictionary \facesDict{}, we append the simplex \simplex{}\ to the value $\facesDict{}[\face{}]$, see Line \ref{line:assignTwoDictionary}. Now we have created a dictionary \facesDict{}\ that contains all the faces of the mesh \mesh{}. Thus, we can determine if a face \face{}\ is a boundary face or an inner face checking how many values has the sorted face \face{}\ inside \facesDict{}, see Lines from \ref{line:typeOneFaceDictionary} to \ref{line:typeTwoFaceDictionary}. If some face appears more than two times, the algorithm stops because a face cannot be shared by more than two simplices. Finally, we return as output the dictionaries \innerDict{}\ and \boundaryDict{}. We recall that we sort lexicographically all the faces of \mesh{}\ to create the keys of the dictionaries \innerDict{}\ and \boundaryDict{}. The order of those faces inside the simplices that contain them could be different from the lexicographical order used to define the keys of the dictionaries.

\begin{algorithm}[t!]
	\caption{Checking of conformity}
	\label{alg:isConformal}
	\begin{algorithmic}[1]
		\Require{\type{Mesh} \mesh{1}, \type{Conformal Mesh} \mesh{0}}
		\Ensure{\type{Bool} \texttt{isConformal}}
		\Function{isMeshConformal}{\mesh{1}, \mesh{0}}
		\State $\innerDict{0}, \boundaryDict{0} = \Call{getFaces}{\mesh{0}}$
		\label{line:getBoundaryDict1}
		\State $\innerDict{1}, \boundaryDict{1} = \Call{getFaces}{\mesh{1}}$
		\label{line:getBoundaryDict2}
		\For{$\face{1}$ \textbf{in} $\texttt{keys}(\boundaryDict{1})$}
		\State $(\vertex{i_{0}}, \vertex{i_{1}}, \ldots , \vertex{i_{n-1}}) = \Call{getFathersVertices}{\face{1}}$
		\label{line:verticesFatherConformal}
		\If{$(\vertex{i_{0}}, \vertex{i_{1}}, \ldots , \vertex{i_{n-1}})$ \textbf{not in} $\texttt{keys}(\boundaryDict{0})$}
		\State \Return \texttt{false}
		\EndIf
		\EndFor
		\State \Return \texttt{\textbf{true}}
		\EndFunction
	\end{algorithmic}
\end{algorithm}
After the definition of the dictionaries \innerDict{}\ and \boundaryDict{}, we propose an algorithm to check the conformity of the mesh, checking that the boundary faces of \mesh{1}\ are contained inside the boundary faces of \mesh{0}. The procedure is depicted in Algorithm \ref{alg:isConformal}. Thus, for a given conformal mesh \mesh{0}\ and a mesh \mesh{1}, the function that check if a mesh is conformal, returns a boolean value depending if the boundary of \mesh{1}\ is contained, or not, inside the boundary of \mesh{0}. That is, if \mesh{1}\ is conformal to \mesh{0}. First, we compute the boundary dictionaries of \mesh{0}\ and \mesh{1}\, respectively, see Lines \ref{line:getBoundaryDict1} and \ref{line:getBoundaryDict2}. When we compute the dictionaries, we are checking that there are no faces that appear more than two times. After that, for each sorted boundary face \face{1}\ of \mesh{1}, we obtain its parent face \face{0}\ of \mesh{0}. To do it so, we first collect in a list all the indices in all the multi-indices of \face{1}. Then, we extract the unique indices of this list. We should obtain a list of length $n$, $(\vertex{i_{0}}, \vertex{i_{1}}, \ldots , \vertex{i_{n-1}})$. This list of vertices is converted to a face \face{0}. Since all the vertices are of length one, we check if the face belongs to the boundary of \mesh{0}\ by checking the dictionary of boundary faces \boundaryDict{0}. Otherwise, the boundary face \face{1}\ is not contained inside a boundary face of \mesh{0}, and the mesh is not conformal.

\subsection{Reflectivity check}
\label{sec:reflectivity}

\begin{algorithm}[t!]
	\caption{Checking of reflected neighbors}
	\label{alg:isReflected}
	\begin{algorithmic}[1]
		\Require{\type{ConformalMesh} \mesh{}}
		\Ensure{\type{Bool}}
		\Function{isReflected}{\mesh{}}
		\State $\innerDict{}, \boundaryDict{} =$ \Call{getFaces}{\mesh{}}
		\label{line:getInnerFacesDictionary}
		\For{$\face{} \in \texttt{keys}(\innerDict{})$}
		\State $(\vertex{0}, \vertex{1}, \ldots , \vertex{n-1}) = \face{}$
		\State $\simplex{1}, \simplex{2} = \innerDict{}[\face{}]$
		\State $(\vertex{1,0}, \vertex{1,1}, \ldots , \vertex{1,n}) = \simplex{1}$
		\State $(\vertex{2,0}, \vertex{2,1}, \ldots , \vertex{2,n}) = \simplex{2}$
		\State $\wertex{1} = $ \Call{oppositeVertex}{\simplex{1}, \face{}}
		\label{line:wertex1}
		\State $\wertex{2} = $ \Call{oppositeVertex}{\simplex{2}, \face{}}
		\label{line:wertex2}
		\State $\face{1} = $ \Call{remove}{\simplex{1}, \wertex{1}}
		\label{line:face1}
		\State $\face{2} = $ \Call{remove}{\simplex{2}, \wertex{2}}
		\label{line:face2}
		\State $(\vertex{i_{0}}, \vertex{i_{1}}, \ldots , \vertex{i_{n-1}}) = \face{1}$
		\State $(\vertex{j_{0}}, \vertex{j_{1}}, \ldots , \vertex{j_{n-1}}) = \face{2}$
		\For{$k = 0, 1, \ldots, n-1$}
		\label{line:loopReflected1}
		\If{$v_{i_{k}} \neq v_{j_{k}}$}
		\State \Return \texttt{false}
		\EndIf
		\EndFor
		\label{line:loopReflected2}
		\EndFor
		\State \Return \texttt{true}
		\EndFunction
	\end{algorithmic}
\end{algorithm}
To check that \meshQ{n}{}\ is reflected, we have to check that all neighboring simplices are reflected neighbors. To do that, we use Algorithm \ref{alg:isReflected}. First, we obtain the dictionary \facesDict{} of the interior faces of \meshQ{n}{}, see Line \ref{line:getInnerFacesDictionary}. For each inner face $\face{} \in \meshQ{n}{}$, the operation $\innerDict{}[\face{}]$ return the two neighboring simplices \simplex{1}\ and \simplex{2} that share \face{}. The face $\face{} = (\vertex{0}, \ldots , \vertex{n-1})$, with $v_{0} < v_{1} < \ldots < v_{n}$, has lexicographic order in \facesDict{}, but it may have a non-shared order inside \simplex{1}\ and \simplex{2}. After obtaining the \simplex{1}\ and \simplex{2}, we obtain the different vertices \wertex{1}\ and \wertex{2}\ between the simplices \simplex{1}\ and \simplex{2}\ and the face \face{}, respectively, see Lines \ref{line:wertex1} and \ref{line:wertex2}. Then, we obtain the faces \face{1}\ and \face{2}\ inside \simplex{1}\ and \simplex{2}\ with the order in which they appear inside each simplex, see Lines \ref{line:face1} and \ref{line:face2}. We recall that \face{1}, \face{2}, and \face{}\ are composed of the same vertices, nevertheless, the vertices may appear in a different order. What we want to check is that \face{1}\ and \face{2}\ have the same order of vertices. To this end, we loop on the faces \face{1}\ and \face{2}, see Lines \ref{line:loopReflected1}--\ref{line:loopReflected2}, checking that they have the same vertices and in the same order.

\subsection{Mesh renumbering}
\label{sec:renumber}

After obtaining a locally refined mesh, we renumber the resulting multi-ids to have new multi-ids of length one and thus,  reduce the memory usage. In Line \ref{line:renumberMesh} of local refinement algorithm, see Algorithm \ref{alg:LocalRefineAlgorithm}, we perform a renumber of the vertices of the generated mesh. Our algorithm identifies the vertices of the mesh with  multi-ids, see Section \ref{sec:definitions}. At the end of the local refinement algorithm, we perform a renumber of the multi-ids of \mesh{k+1}\ to multi-ids of length one. To that purpose, we sort lexicographically all the multi-ids of \mesh{k+1}\ obtaining a sorted list of multi-ids. Then, we assign to the $i$-th multi-id of the sorted list the new multi-id of length one $[i]$. We have to perform that renumber of multi-ids in all the data structures where the multi-ids appear.

\end{document}